\documentclass{report}

\usepackage{graphicx} 
\usepackage{subfigure} 
\usepackage{lipsum}

\usepackage{natbib}

\usepackage{algpseudocode}
\usepackage{algorithm}
\usepackage{enumitem}
\usepackage{color}
\usepackage{url}
\usepackage{amsmath}
\usepackage{amssymb}
\usepackage{xspace}
\usepackage{amsfonts}
\usepackage{booktabs}
\usepackage{siunitx}
\usepackage{tabularx}
\usepackage[T1]{fontenc}
\usepackage[utf8]{inputenc}
\usepackage{graphicx}
\usepackage[flushleft]{threeparttable}

\usepackage{array}
\usepackage{calc}

\newcommand{\cov}{covariates\xspace}

\newcommand{\sti}{stopout\xspace}
\newcommand{\selfself}{self-proposed, self-extracted\xspace}
\newcommand{\crowdself}{crowd-proposed, self-extracted\xspace}

\newcommand{\lag}{\textit{lag}\xspace}
\newcommand{\neither}{passive collaborator\xspace}
\newcommand{\wiki}{wiki contributor\xspace}
\newcommand{\forum}{forum contributor\xspace}
\newcommand{\both}{fully collaborative\xspace}
\newcommand{\corr}{\textit{correlative}\xspace}
\newcommand{\pred}{\textit{predictive}\xspace}
\newcommand{\beha}{\textit{behavioral}\xspace}
\newcommand{\nbeha}{\textit{non-behavioral}\xspace}
\newcommand{\timev}{\textit{time varying}\xspace}
\newcommand{\timeiv}{\textit{time-invariant}\xspace}

\newcommand{\x}[1]{$x_{#1}$\xspace}

\newcommand{\tten}{\textit{average pre deadline submission time}\xspace}
\newcommand{\tnine}{\textit{correct submissions percent}\xspace}

\newcommand{\tseven}{\textit{lab grade over time}\xspace}
\newcommand{\tsix}{\textit{lab grade}\xspace}
\newcommand{\tfive}{\textit{pset grade over time}\xspace}
\newcommand{\tfour}{\textit{pset grade}\xspace}
\newcommand{\tthree}{\textit{average number of submissions in percent}\xspace}

\newcommand{\feleven}{\textit{submissions per correct problem}\xspace}
\newcommand{\ffive}{\textit{average length of forum post}\xspace}

 \usepackage[accepted]{icml2013}

\begin{document} 

\twocolumn[
\icmltitle{Likely to stop? \\ Predicting Stopout in Massive Open Online Courses}

\icmlauthor{Colin Taylor}{colin\_t@mit.edu}
\icmladdress{Massachusetts Institute of Technology,
           Cambridge, MA 02139 USA}
\icmlauthor{Kalyan Veeramachaneni}{kalyan@csail.mit.edu}
\icmladdress{Massachusetts Institute of Technology,
         Cambridge, MA 02139 USA}
\icmlauthor{Una-May O'Reilly}{unamay@csail.mit.edu}
\icmladdress{Massachusetts Institute of Technology,
           Cambridge, MA 02139 USA}

\icmlkeywords{boring formatting information, machine learning, ICML}

\vskip 0.3in
]

\begin{abstract} 
Understanding why students stopout will help in understanding how students learn in MOOCs. In this report, part of a 3 unit compendium, we describe how we build accurate predictive models of MOOC student stopout. We document a scalable, stopout prediction methodology, end to end, from raw source data to model analysis. We attempted to predict stopout for the Fall 2012 offering of 6.002x. This involved the meticulous and crowd-sourced engineering of over 25 predictive features extracted for thousands of students, the creation of temporal and non-temporal data representations for use in predictive modeling, the derivation of over 10 thousand models with a variety of state-of-the-art machine learning techniques and the analysis of feature importance by examining over 70000 models. We found that stop out prediction is a tractable problem. Our models achieved an AUC (receiver operating characteristic area-under-the-curve) as high as $0.95$ (and generally $0.88$) when predicting one week in advance. Even with more difficult prediction problems, such as predicting stop out at the end of the course with only one weeks' data, the models attained AUCs of $0.7$. 
\end{abstract} 

\section{Introduction}
Massive Open Online Courses (MOOCs) leverage digital technologies to teach advanced topics at scale. MOOC providers such as edX and Coursera boast hundreds of classes developed by top-tier universities. Renowned professors record their lectures, and when needed, use interactive whiteboards to explain concepts. Recordings are delivered all over the world via web servers at no cost to the learner.
Far from compromising the quality of course content, the internet provides a flexible medium for educators to employ new instructional tools. For example, videos enable students to pause, rewind, review difficult concepts and even adjust the speed. In addition, MOOCs allow the learner the flexibility to learn in his or her own time frame. Only in the online medium are short lectures logistically feasible through videos. MOOCs are changing the face of education by providing an alternative to the ``one size fits all",  learning concept employed by hundreds of universities.

The specific layout of each MOOC varies, but most follow a similar format. Content is sectioned into modules, usually using weeks as intervals. Most MOOCs include online lectures (video segments), lecture questions, homework questions, labs, a forum, a Wiki, and exams. Students advance through the material sequentially, access online resources, submit assignments and participate in peer-to-peer interactions (like the forum).

Not surprisingly, MOOCs have attracted the attention of online learners all over the world. The platforms boast impressive numbers of registrants and individuals who complete online course work. For example, MITx offered its first course 6.002x: Circuits and Electronics in the Fall of 2012. 6.002x had 154,763 registrants. Of those, 69,221 students looked at the first problem set, and 26,349 earned at least one point. 9,318 students passed the midterm and 5,800 students got a passing score on the final exam. Finally, after completing 15 weeks of study, 7,157 registrants earned the first certificate awarded by MITx, showing they had successfully completed 6.002x. For perspective, approximately 100 students take the same course each year at MIT. It would have taken over 70 years of on-campus education to grant the same number of 6.002x certificates that were earned in a single year online.

While the completion rates are impressive when compared to in-class capacity, they are still low relative to the number of people who registered, completed certain parts of the course or spent a considerable amount of time on the course. To illustrate, in the above scenario approximately 17\% attempted and got at least one point on the first problem set. The percentage of students who passed the midterm drops to just 6\%, and certificate earners dwindles at just under 5\%. 94\% of registrants did not make it past the midterm.

How do we explain the 96\% \sti \footnote{We use stopout as synonymous with dropout and we refer to its opposite as persistence} rate from course start to course finish? Analyzing completion rates goes hand in hand with understanding student behavior. One MOOC research camp advocates analyzing student usage patterns-- resources used, homework responses, forum and Wiki participation -- to improve the online learning experience thereby increasing completion rates. Other researchers question the feasibility of analyzing completion rates altogether because the online student body is unpredictable. For example, some students register online because it is free and available with little or no intention of finishing. Some students who leave may lack motivation, or could leave due to personal reasons completely unrelated to MOOCs. As a result, interpreting completion rates is not a straightforward exercise. However, we believe that if we are to fully understand how students learn in MOOCs, we need to better understand why students \sti. Building accurate predictive models is the first step in this undertaking.

\noindent \textbf{Why predict stopout?}
There are a number of reasons to predict stopout.
\begin{description}
\vspace{-2mm}
\item \textbf{Interventions}: Stopout prediction in advance allows us to design interventions that would increase engagement, provide motivation and eventually prevent \sti.
\vspace{-2mm}
\item \textbf{Identifying intentions}:  Certain special cases of \sti prediction allow us to delineate the student intentions in taking the MOOC. For example, the cohort for whom we are able to predict \sti accurately based on just their first week behavior could imply that the method and manner in which the course was designed or handled had no effect on the learner's decision to \sti. 
\vspace{-4mm}
\item \textbf{Model analysis}: Analysis of accurate statistical models that have maximum \sti prediction accuracy can yield insights as to what caused the students to \sti. From  this perspective one can even examine the students for whom the predictions were wrong, \textit{that is}, a very accurate and trustworthy model predicted the student would not \sti but the student did \sti. Here the model error could be due to reasons that are unrelated to the course itself.
\vspace{-3mm}
\end{description}

In a compendium of papers, of which this is first of three, we tackle the challenge of predicting student persistence in MOOCs. Throughout the compendium we focus on the aforementioned course, the Fall 2012 offering of 6.002x: Circuits and Electronics.  We believe a three pronged approach which comprehensively analyzes student interaction data, extracts from the data sophisticated predictive indicators and leverages state-of-the-art models will lead to successful predictions. The compendium presents a comprehensive treatment of predicting \sti which produces and considers complex, multi-layered yet interpretive features and fine tuned modeling. 

We ask whether it is possible for machine learning algorithms, with only a few weeks of data, to accurately predict persistence. Is it possible to predict, given only the first week of course data, who will complete the last week of the course?  How much history  (or how many weeks' data) is necessary for accurate prediction one or more week ahead?

\subsection{Outline of the compendium}

The compendium is organized into the following papers:
\begin{itemize}
\item In this paper, we describe the stopout prediction problem, and present a number of discriminatory models we built starting with Logistic regression and moving to Support Vector Machines, Deep Belief networks and decision trees. We also present a summary of which features/variables played a role in gaining accurate predictions. 

\item In ``\textit{Towards Feature Engineering at Scale for Data from Massive Open Online Courses"} we present how we approached the problem of constructing interpretive features from a time series of click stream events. We present the list of features we have extracted to create the predictive models \cite{featMOOC}.

\item In ``\textit{Exploring Hidden Markov Models for modeling online learning behavior"} we outline a temporal modeling technique called Hidden Markov models and present the results when these models are used to make predictions. We also present a stacked model using both techniques (HMMs and Logisitic regression) and present overview of our findings \cite{hmmMOOC}.
\end{itemize}
\paragraph{Compendium contributions}: The most fundamental contribution of this compendium is the design, development and demonstration of a stopout prediction \textbf{methodology}, end to end, from raw source data to model analysis. The methodology is painstakingly meticulous about every detail of data preparation, feature engineering, model evaluation and outcome analysis. Our goal with such thoroughness is to advance the state of research into stopout from its current position and document a methodology that is reproducible and scalable. We will next generalize this methodology to a number of additional edX and Coursera courses and report the successes and limitations. In addition, the methodology and software will shortly be released to interested educational researchers.

\subsection{Our contributions through this paper}
This paper makes the following contributions:
\vspace{-3mm}
\begin{itemize}
\item We successfully predict stopout for the Fall 2012 offering of 6.002x. The major findings of the predictive models are presented in ~\ref{sect:findings}. 
\item We extract 27 sophisticated, interpretive features which combine student usage patterns from different data sources. This included leveraging the collective brain-power of the crowd. These are presented in ~\ref{sect:feats}. 
\item We utilize these features to create a series of temporal and non-temporal feature-sets for use in predictive modeling.
\item We create over 10,000 comprehensive, predictive models using a variety of state-of-the-art techniques.
\item We demonstrate that with only a few weeks of data, machine learning techniques can predict persistence remarkably well. For example we were able to achieve an area under the curve of the receiver operating characteristic of 0.71, given only one week of data, while predicting student persistence in the last week of the course. Given more data, some of the models reached an AUC of 0.95. We present these and other results in ~\ref{sect:results}. 
\item We build and demonstrate a scalable, distributed, modular and reusable framework to accomplish these steps iteratively.
\end{itemize}

The rest of the paper is organized as follows. Section~\ref{sect:dataorg} presents the details of the data we use, its organization, features/variables we extracted/operationalized for modeling. Section~\ref{sect:pred} presents the definition of the prediction problem and the different assumptions we make in defining the problem. Section~\ref{sect:logreg} presents the predictive modeling technique - \textit{logistic regression} and the results we achieved for all 91 prediction problems. Section~\ref{sect:mulitple_classifiers} presents the details of how we employed multiple predictive modeling techniques. Section~\ref{sect:related} presents the related work both prior to MOOCs and MOOCs. Section~\ref{sect:findings} presents the research findings relevant to this paper. Section~\ref{sect:general} presents our reflection for the entire compendium.

\section{Data organization into MOOCdb}\label{sect:dataorg}
As previously mentioned, we focused on the Fall 2012 offering of 6.002x: Circuits and Electronics. edX provided the following raw data from the 6.002x course:
\begin{itemize}
\item A dump of click-stream data from learner-browser and edX-server tracking logs in JSON format. For instance, every page visited by every learner was stored as a server-side JSON (JavaScript Object Notation) event.
\item Forum posts, edits, comments and replies stored in a MongoDB collection. Note that passive forum data, such as how many views a thread received was not stored here and had to be inferred from the click-stream data.
\item Wiki revisions stored in a MongoDB collection. Again, passive views of the Wiki must be inferred from the click-stream data.
\item A dump of the MySQL production database containing learner state information. For example, the database contained his/her final answer to a problem, along with its correctness. Note that the history of his submissions must be inferred from the click-stream data.
\item An XML file of the course calendar which included information like the release of content and the assignment deadlines.
\end{itemize}
\begin{figure*}[ht!]
  \centering
    \includegraphics[width=0.6\textwidth]{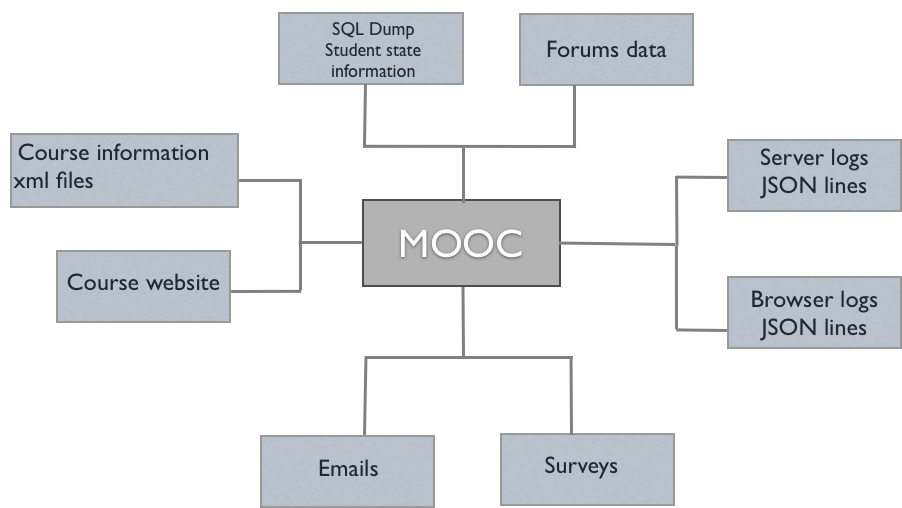}
     \caption{Multiple data sources received from edX with their corresponding formats}\label{fig:data_layout}
\end{figure*}
Figure \ref{fig:data_layout} summarizes the raw data received. This data included:
\begin{itemize}
\item 154,763 registered learners
\item 17.8 million submission events
\item 132.3 million navigational events \footnote{We received more navigational events, but only 132.3 million were well formed enough to be reliably considered for this compendium. }
\item $\sim$90,000 forum posts
\end{itemize}

To analyze this data at scale, as well as write reusable analysis scripts, we first organized the data into a schema designed to capture pertinent information. The resulting database schema, MOOCdb, is designed to capture MOOC data across platforms thereby promoting collaboration among MOOC researchers. MOOCdb utilizes a large series of  scripts to pipe the 6.002x raw data into a standardized schema. More about MOOCdb can be found in the MOOCdb Tech report, but the details are outside the scope of this compendium \cite{tr}.

\begin{figure*}[ht!]
  \centering
    \includegraphics[width=0.6\textwidth]{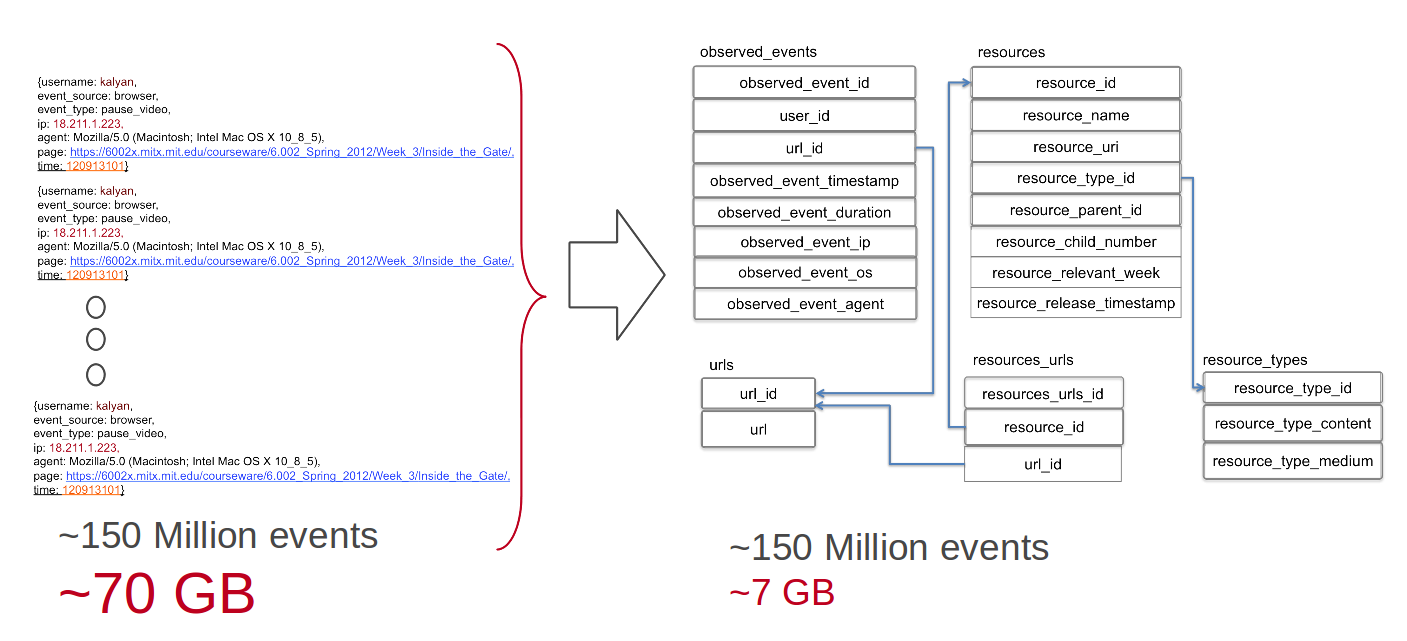}
     \caption{Piping data into MOOCdb}\label{fig:data_reduction}
\end{figure*}

Through the labor intensive process of piping the raw data into a schematized database, we were able to significantly reduce the data size in terms of disk space. The original $\sim$70GB of raw data was reduced to a $\sim$7GB MOOCdb through schema normalization. The transformation was crucial in order to load the entire database into RAM enabling prompt queries and feature extractions. Figure~\ref{fig:data_reduction} shows a snapshot of the original JSON transactional data transformed into a normalized schema.

\section{Prediction problem definition and assumptions}\label{sect:pred}
We made several assumptions to more precisely define the \sti prediction problem and interpret the data. These assumptions include time-slice delineation and defining persistence (\sti) as the event we attempt to predict.

\subsection{Time-slice delineation}
Temporal prediction of a future event requires us to assemble explanatory variables along a time axis. This axis is subdivided to express the time-varying behavior of variables so they can be used for explanatory purposes. In 6.002x, course content was assigned and due on a weekly basis, where each week corresponded to a module. Owing to the regular modular structure, we decided to define time slices as weekly units. Time slices started the first week in which course content was offered, and ended in the fifteenth week, after the final exam had closed.
\begin{figure*}[ht!]
  \centering
    \includegraphics[width=0.8\textwidth]{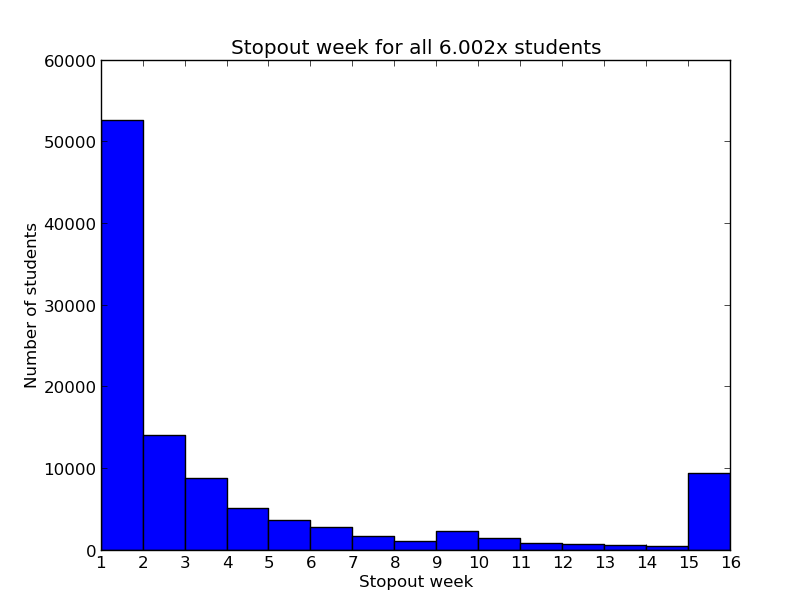}
      \caption{Stopout week distribution}\label{fig:dropout_week}
\end{figure*}
\subsection{Stopout definition}
The next question we had to address was our definition of stopout. We considered defining it by the learner's last interaction in the course, regardless of the nature of the interaction. This is the approach taken by Balakrishnan in his \sti analysis \cite{balakrishnan2013predicting}. However, Balakrishnan's definition yields noisy results because it gives equal weight to a passive interaction (viewing a lecture, accessing an assignment, viewing a Wiki etc) as it does to a pro-active interaction (submitting a problem, midterm, assignment etc). A learner could stop submitting assignments in the course after week 2, but continue to access the course pages and not be considered stopped out. Instead, we define the stopout point as the time slice (week) a learner fails to submit any further assignments or exercise problems. To illustrate, if a learner submits his/her last assignment in the third module, he/she is considered to have stopped out at week four. A submission (or attempt) is a submission of any problem type (Homework, lab, exam etc.), as defined in MOOCdb. This definition narrows the research to learners who consistently participate in the course by submitting assignments. Using this definition for \sti we extracted the week number when each learner in the cohort stopped out.

Figure~\ref{fig:dropout_week} shows the distribution of \sti week for all 105,622 learners who ever accessed the course. Of these, 52,683 learners stopped out on week one. These learners never submitted an assignment, and are never considered in the rest of our analysis. Another large learner drop off point is in week 15, the last week of the course. Many of these learners actually finished the course, but did so by submitting their final exam in week 14. This nuance presented itself because learners had a range of time to start the final exam, and this range actually overlapped between weeks 14 and 15. Due to the nature of the final exam time range, we never attempt to predict week 15, and consider week 14 as the final week.

\begin{figure*}[!ht]
   \centering
    \includegraphics[width=0.8\textwidth]{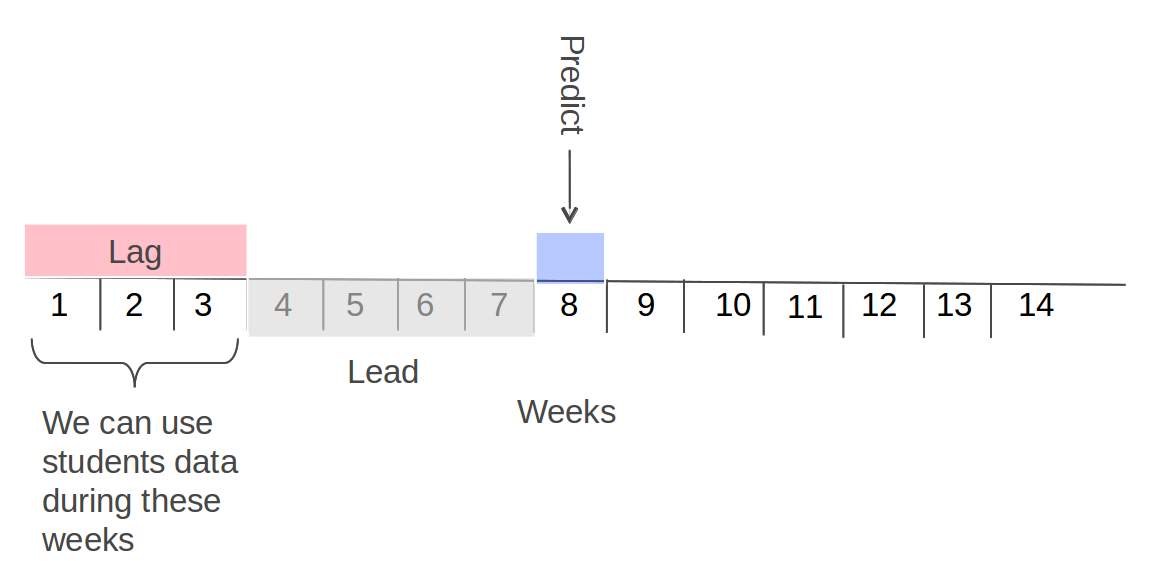}
    \caption{Diagram of the learners' weeks data used in a lead 5, lag 3 prediction problem}\label{fig:leadlag}
\end{figure*}

\subsection{Lead and Lag}
Lead represents how many weeks in advance to predict \sti. We assign the \sti label (\x{1}, 0 for \sti or 1 for persisted) of the lead week as the predictive problem label. Lag represents how many weeks of historical variables will be used to classify. For example, if we use a lead of 5 and a lag of 3, we would take the first 3 weeks of data to predict 5 weeks ahead. Thus, each training data point is a learner's feature values for weeks 1, 2 and 3 as features. The binary \sti value for week 8 becomes the label. Figure~\ref{fig:leadlag} shows a diagram of this scenario.

We are careful not to use learners' stopped out week's features as input to our models. In other words, if a learner has stopped out in week 1, 2 or 3, we do not use this learner as a data point. Including stopped out learner data makes the classification problem too easy as the model will learn that a stopped out learner never returns (by our \sti definition).

To illustrate the predictive model's potential application, we will use a realistic scenario. The model user, likely an instructor or platform provider, could use the data from week 1 to week $i$ (current week) to make predictions. The model will predict existing learner \sti during weeks $i+1$ to $14$. For example, Figure~\ref{fig:leadlag} shows one such prediction problem. In this case the user, currently at the end of week 3, is attempting to predict \sti for the 8th week. 

\begin{figure*}[ht!]
    \centering
    \includegraphics[width=0.7\textwidth]{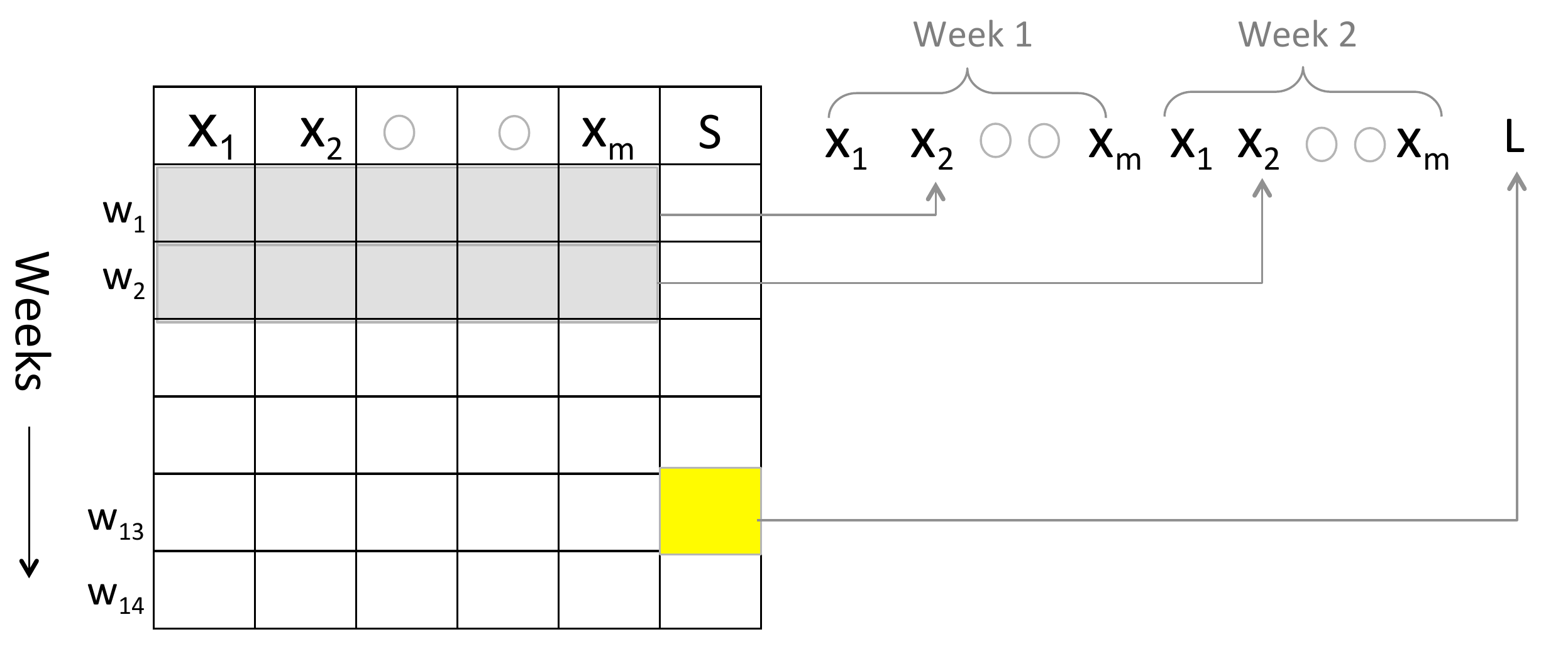}
    \caption{Diagram of the flattening process. In this case two weeks of data is used to predict week 13. This prediction problem corresponds to a lead of 11, and a lag of 2.}\label{fig:flattening}
\end{figure*}

\noindent \textbf{Multiple prediction problems}\label{para:multiple_prediction}
Under this definition 91 individual prediction problems exist. For any given week $i$ there are $14-i$ number of prediction problems. Each prediction problem becomes an independent modeling problem which requires a discriminative model. To build discriminative models we utilize a common approach of flattening out the data, that is forming the \cov for the discriminative model by assembling the features from different learner-weeks as separate variables. This process is shown in Figure~\ref{fig:flattening}. The example uses data from weeks 1 and 2 (lag of 2) and attempts to predict the \sti for week 13 (lead of 11). 

\begin{figure*}[!ht]
  \centering
    \includegraphics[width=0.8\textwidth]{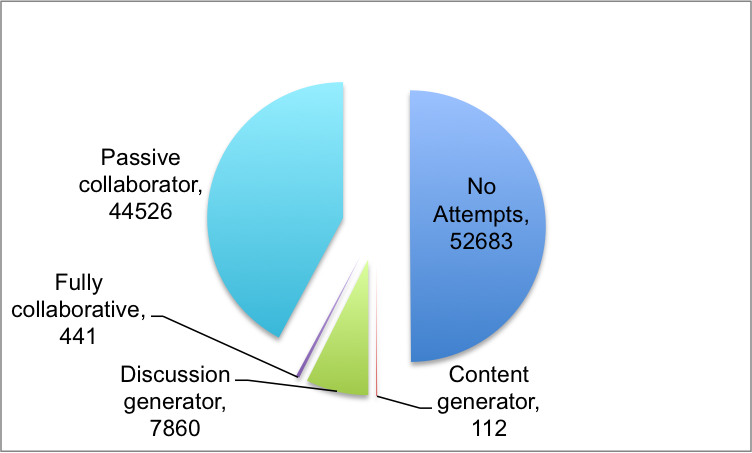}
      \caption{Chart of the relative sizes of our cohorts}\label{fig:cohortsplit}
\end{figure*}
For all of the ensuing modeling and analysis, we treated and reported on each of the cohort datasets independently.
\begin{figure*}[ht!]
  \centering
    \includegraphics[width=0.5\textwidth]{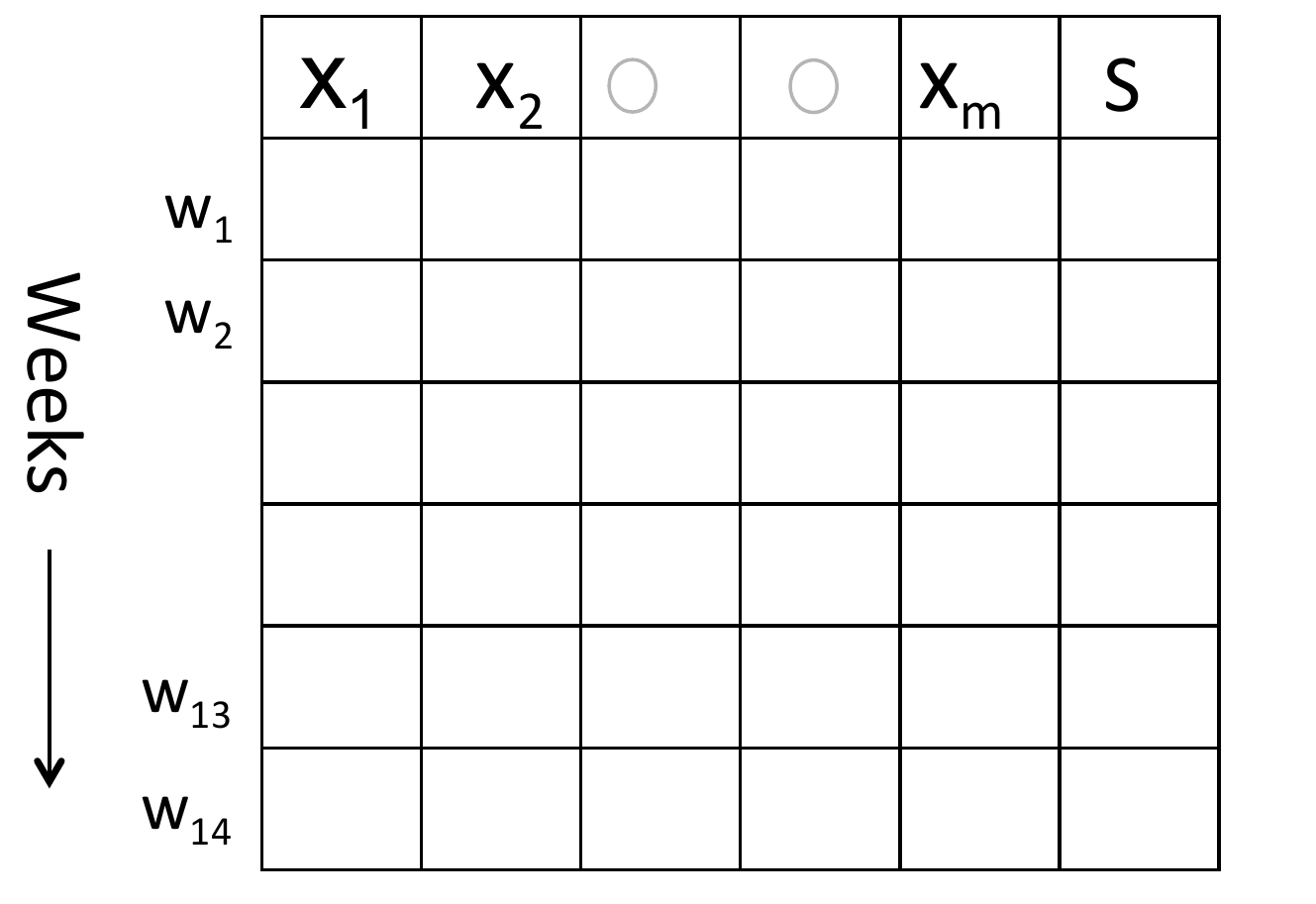}
     \caption{The feature matrix, which captures each feature value for each week. Each student has such a matrix.}\label{fig:student_weekly_features}
\end{figure*}

\subsection{Partitioning learners into cohorts}
Rather than treat all learners uniformly, we decided to build predictive models for different types of learners. With this in mind we divided the learners into cohorts as a rough surrogate variable for their commitment to the course. We chose four cohorts based on the learner's collaborative activity throughout the course. More specifically, we divided learners based on whether or not they participated in the class forum or helped edit the class Wiki pages. The four types of learners are:
\vspace{-4mm}
\begin{itemize}
\item \neither - these learners never actively participated in either the forum or the Wiki. They are named passive because they passively viewed, but did not contribute to, resources.
\vspace{-1mm}
\item \wiki - these learners actively participated in the Wiki by generating Wiki content through their edits, but never actively posted in the forum.
\vspace{-1mm}
\item \forum - these learners actively posted in the forum, but never actively participated in the class Wiki.
\vspace{-1mm}
\item \both - these learners actively participated by generating Wiki content and by posting in the forum
\end{itemize}
\vspace{-4mm}
From the combined dataset of 52,939 participating learners, we assigned each learner into one of the four types. The chart ~\ref{fig:cohortsplit} summarizes the sizes of the cohort datasets.

\subsection{Features per learner}\label{sect:feats}
We extracted 27 interpretive features on a per-learner basis. These are the features we use to build a model. We describe the process of feature engineering at length in \cite{featMOOC}. In this paper for the sake of brevity, we only list the features and their brief descriptions in the two tables below. For more details about how we came up with these features and how specifically these features were operationalized we refer the readers to \cite{featMOOC}
\setlength{\arrayrulewidth}{0.5pt}
\setlength{\tabcolsep}{10pt}
\renewcommand{\arraystretch}{1.5}

\newlength{\thickline}
\setlength{\thickline}{1pt}
\makeatletter
\def\hlinex{%
  \noalign{\ifnum0=`}\fi\hrule \@height \thickline \futurelet
   \reserved@a\@xhline}
\makeatother

\newlength{\threecoltabwid}
\setlength{\threecoltabwid}{\textwidth - \tabcolsep * 2 * 3}

\begin{table*}[htp]
\centering
\begin{threeparttable}
 \caption{List of \selfself \cov}\label{table:self_proposed_self_extracted}
\medskip
  \begin{tabular*}{\textwidth}{ >{\centering\arraybackslash}p{0.05\threecoltabwid} >{\raggedright\arraybackslash}p{0.25\threecoltabwid} >{\raggedright\arraybackslash}p{0.7\threecoltabwid}}
\hlinex
& \textbf{Name}  & \textbf{Definition}  \\ \hlinex
\x{1}  & stopout & Whether the student has stopped out or not  \\ \hline
*\x{2} & total duration& Total time spent on all resources  \\ \hline
\x{3}  & number forum posts  & Number of forum posts\\ \hline
\x{4}  & number wiki edits& Number of wiki edits\\ \hline
*\x{5} & average length forum post& Average length of forum posts\\ \hline 
*\x{6}  & number distinct problems submitted & Number of distinct problems attempted \\ \hline 
*\x{7} & number submissions  & Number of submissions \tnote{1}\\ \hline
\x{8}  & number distinct problems correct & Number of distinct correct problems \\ \hline 
\x{9} & average number submissions & Average number of submissions per problem (\x{7} / \x{6})\\ \hline 
\x{10} & observed event duration per correct problem  & Ratio of total time spent to number of distinct correct problems (\x{2} / \x{8}). This is the inverse of the percent of problems correct \\ \hline 
\x{11} & submissions per correct problem  & Ratio of number of problems attempted to number of distinct correct problems (\x{6} / \x{8}) \\ \hline
\x{12} & average time to solve problem & Average time between first and last problem submissions for each problem (average(max(submission.timestamp) - min(submission.timestamp) for each problem in a week) )\\ \hline
*\x{13} & observed event variance& Variance of a student's observed event timestamps    \\ \hline
\x{14} & number collaborations& Total number of collaborations (\x{3} + \x{4})   \\ \hline
\x{15} & max observed event duration & Duration of longest observed event  \\ \hline
*\x{16} & total lecture duration& Total time spent on lecture resources \\ \hline
*\x{17} & total book duration & Total time spent on book resources  \\ \hline
*\x{18} & total wiki duration & Total time spent on wiki resources  \\ \hlinex
  \end{tabular*}
 \medskip
\begin{tablenotes}
\footnotesize
\item[1] In our terminology, a submission corresponds to a problem attempt. In 6.002x, students could submit multiple times to a single problem. We therefore differentiate between problems and submissions.
\end{tablenotes}
\end{threeparttable}
\end{table*}

\setlength{\arrayrulewidth}{0.5pt}
\setlength{\tabcolsep}{10pt}
\renewcommand{\arraystretch}{1.5}
\begin{table*}[htp]
 \centering
 \caption{List of \crowdself \cov}\label{table:crowd_proposed_self_extracted}
 \medskip
\begin{tabular*}{\textwidth}{ >{\centering\arraybackslash}p{0.05\threecoltabwid} >{\raggedright\arraybackslash}p{0.35\threecoltabwid} >{\raggedright\arraybackslash}p{0.6\threecoltabwid}}
 \hlinex
   & \textbf{Name}      & \textbf{Definition}          \\ \hlinex
 $x_{201}$ & number forum responses & Number of forum responses      \\ \hline
 *$x_{202}$ & average number of submissions percentile & A student's average number of submissions (feature 9) as compared with other students that week as a percentile  \\ \hline
 *$x_{203}$ & average number of submissions percent & A student's average number of submissions (feature 9) as a percent of the maximum average number of submissions that week             \\ \hline
 *$x_{204}$ & pset grade    & Number of the week's homework problems answered correctly / number of that week's homework problems                \\ \hline
 $x_{205}$ & pset grade over time   & Difference in grade between current pset grade and average of student's past pset grade                   \\ \hline
 *$x_{206}$ & lab grade    & Number of the week's lab problems answered correctly / number of that week's lab  problems                \\ \hline
 $x_{207}$ & lab grade over time  & Difference in grade between current lab grade and average of student's past lab grade                   \\ \hline
 $x_{208}$ & number submissions correct  & Number of correct submisions      \\ \hline
 $x_{209}$ & correct submissions percent  & Percentage of the total submissions that were correct ($x_{208}$ / $x_{7}$)     \\ \hline
 *$x_{210}$ & average predeadline submission time & Average time between a problem submission and problem due date over each submission that week               \\ \hlinex
\end{tabular*}

\end{table*}
\section{Logistic Regression}\label{sect:logreg}

\begin{figure}[ht!]
  \centering
    \includegraphics[width=0.45\textwidth]{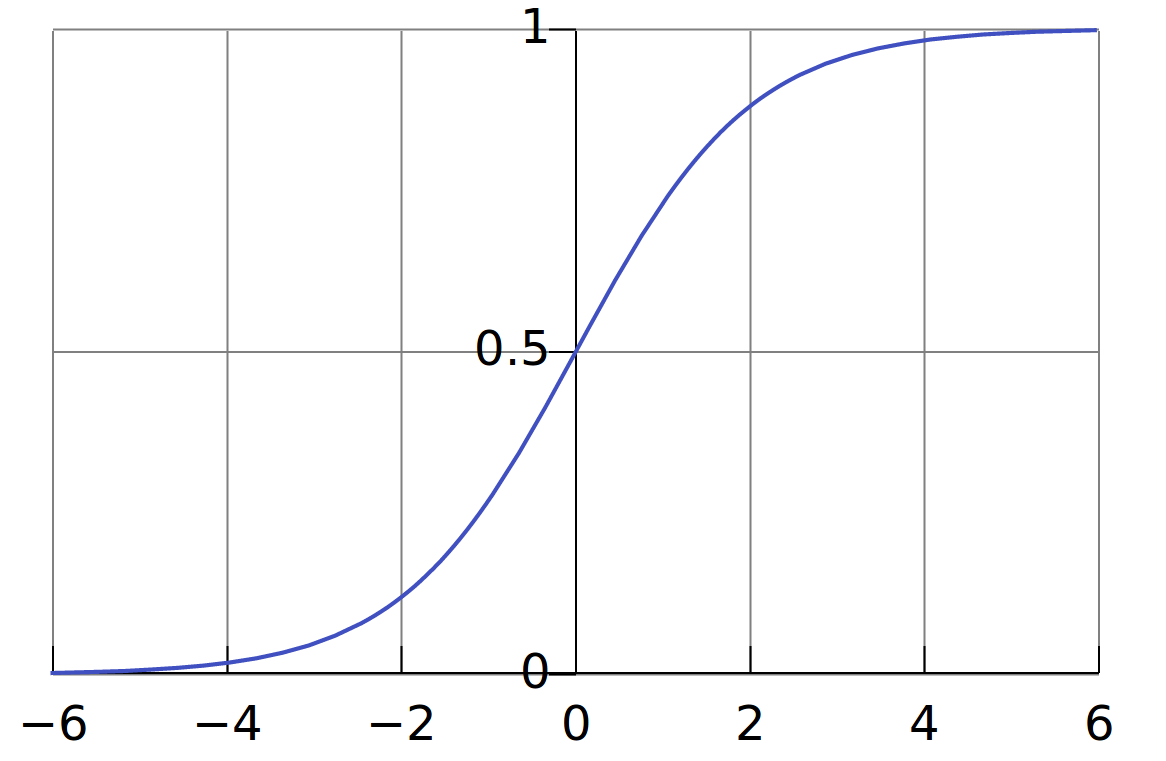}
     \caption{The logit (aka logistic or sigmoid) function. The logit equation is $ y = \frac{1}{1 + e ^{-x}}$. The range of the function is between 0 and 1.}\label{fig:logit}
\end{figure}

Logistic regression is a commonly used binary predictive model. It calculates a weighted average of a set of variables, submitted as \cov, as an input to the \textit{logit} function. Thus, the input to the \textit{logit} function, z, takes the following form:
\begin{equation}
 z = \beta_0 + \beta_1 * x_1 + \beta_2 * x_2 + ... \beta_m * x_m  
 \end{equation}
Here, $\beta_1$ to $\beta_m$ are the coefficients for the feature values, $x_1$ to $x_m$. $\beta_0$ is a constant. The \textit{logit} function, given by, 
\begin{equation}\label{eq:logit}
 y = \frac{1}{1 + e ^{-z}}
 \end{equation}
takes the shape as shown in figure \ref{fig:logit}. Note that the function's range is between 0 and 1, which is optimal for probability. Also note that it tends to `smooth out' at extreme input value, as the range is capped. 

For a binary classification problem, such as ours, the output of the \textit{logit} function becomes the estimated probability of a positive training example. These feature weights, or coefficients, are similar to the coefficients in linear regression. The difference is that the output ranges between 0 and 1 due to the logit function, rather than an arbitrary range for linear regression.

\subsection{Learning}
The objective of training a logistic regression model is to find a set of coefficients well suited to fit the data. For the binary classification problem,  as noted before, training involves passing a set of \cov and a corresponding binary label associated with the \cov. After training a model, the predicted probability, or the output of the \textit{logit} function, should predict higher probabilities for the positive `+1' class examples in the training data and a lower probability for the negative `0' class examples. 

There is no closed form solution to find the optimal coefficients to best fit the training data. As a result, training is usually done iteratively through a technique called maximum likelihood estimation \cite{menard2002applied}. First, a random set of coefficients are chosen. At each iteration, an algorithm such as Newton's method is used to find the gradient between what the coefficients predict and what they should predict, and updates the weights accordingly. The process repeats until the change in the coefficients is sufficiently small. This is called convergence. After running this iterative process over all of the training examples, the coefficients represent the final trained model.

\subsection{Inference and evaluation}
With training in place, the next step is evaluating the classifier's performance. A testing set comprised of untrained \cov and labels evaluates the performance of the model on the test data following the steps below: 
\begin{description}\label{steps}
\item {Step 1:} The logistic function learned and presented in ~\ref{eq:logit} is applied to each data point and the estimated probability of a positive label $y_i$ is produced for each data point in test set.
\item {Step 2:}  A decision rule is applied to determine the class label for each probability estimate $y_i$. The decision rule is given by: 
\begin{equation}\label{eq:decRule}
\hat {L_i} = \left\{ 1, \ if \ y_i \geq \lambda \atop 0, \ if \ y_i < \lambda \right\}
\end{equation}
Given the estimated labels for each data point $\hat{L_i}$ and the true labels $L_i$ we can calculate the confusion matrix, true positives and false positives and thus obtain an operating point on the ROC curve. 
\item {Step 3:} By varying the threshold $\lambda$ in \ref{eq:decRule} the decision rule above we can evaluate multiple points on the ROC curve. We then evaluate the area under the curve and report that as the performance of the classifier on the test data.
\end{description}

\begin{figure}[!ht]
   \centering
    \includegraphics[width=0.45\textwidth]{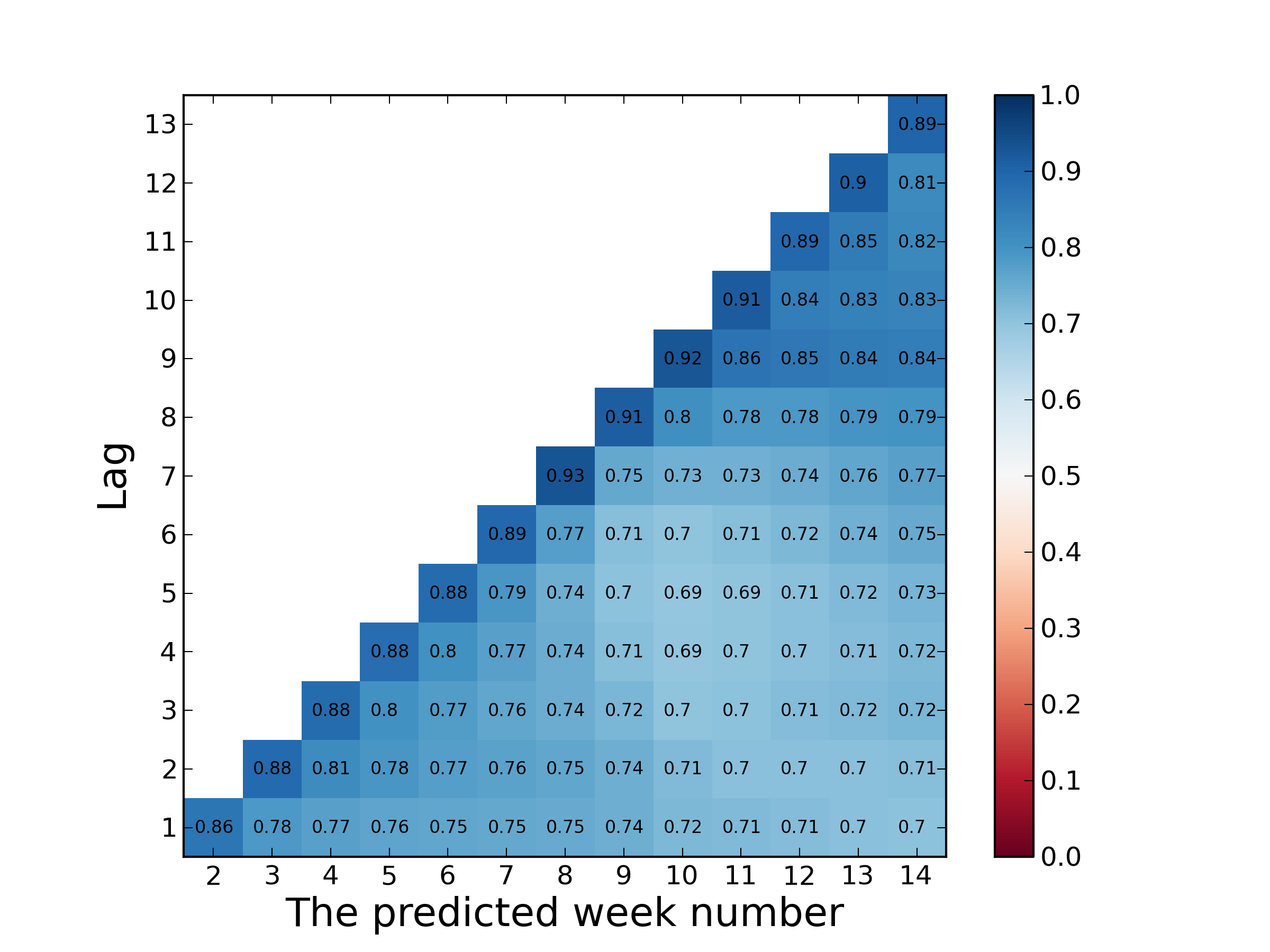}
     \caption{Example heatmap for a logistic regression problem. The heatmap shows how the ROC AUC varied as \lag changed as the target prediction week changed.}\label{fig:logistic_regression_heatmap_no_collab2}
\end{figure}

\begin{paragraph}
{Predictive accuracy heat map} To present the results for multiple prediction problems for different weeks simultaneously, as discussed in Section~\ref{para:multiple_prediction},  we assemble a heat map of a lower right triangular matrix as shown in Figure~\ref{fig:logistic_regression_heatmap_no_collab2}. The number on the x-axis is the week for which predictions are made of that experiment. The y-axis represents the \lag, or the number of weeks of data used to predict. The color represents the area under the curve for the ROC that the model achieved. Note that as the predicted week increases for a given \lag, it is harder to predict. Likewise, as we increase the \lag for a given prediction week, the stopout value becomes easier to predict. This implies that using more historical information enables a better prediction.
\end{paragraph}

\subsection{Attractive properties of logistic regression} 
\begin{itemize}
\item It is relatively simple to understand. 
\item After a model is trained, it provides feature weights, which are useful in assessing the predictive power of features (this will be discussed further in our treatment of the randomized logistic regression model). 
\item It is fast to run. On a single i-7 core machine, for example, running each of the 91 prediction problems on all 4 cohorts took ~25 hours.
\end{itemize}
\begin{figure*}[ht!]
  \centering
    \includegraphics[width=0.8\textwidth]{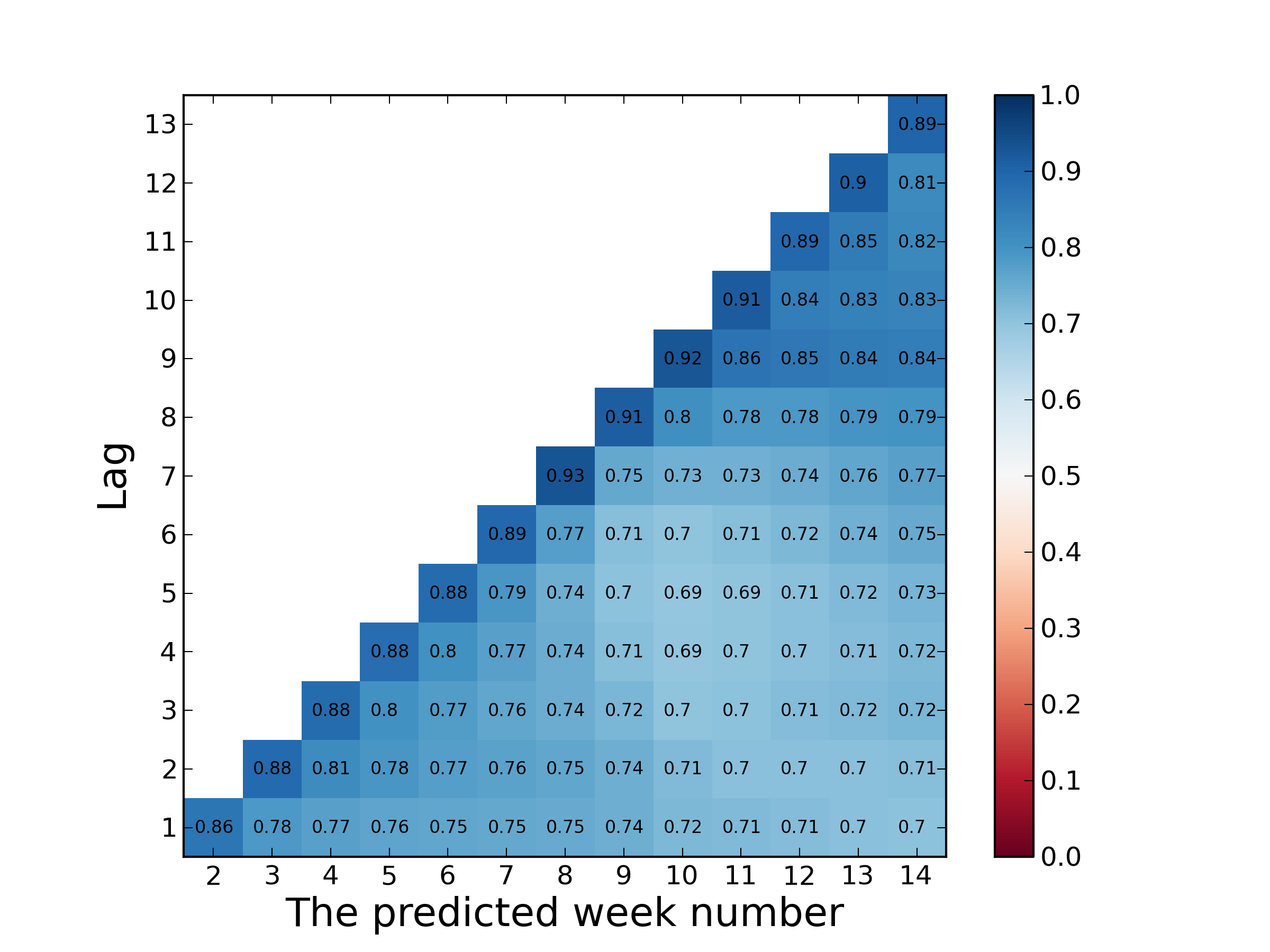}
      \caption{Logistic regression results for the \neither cohort.}\label{fig:logistic_regression_heatmap_no_collab}
\end{figure*}

\begin{figure*}[ht!]
  \centering
    \includegraphics[width=0.8\textwidth]{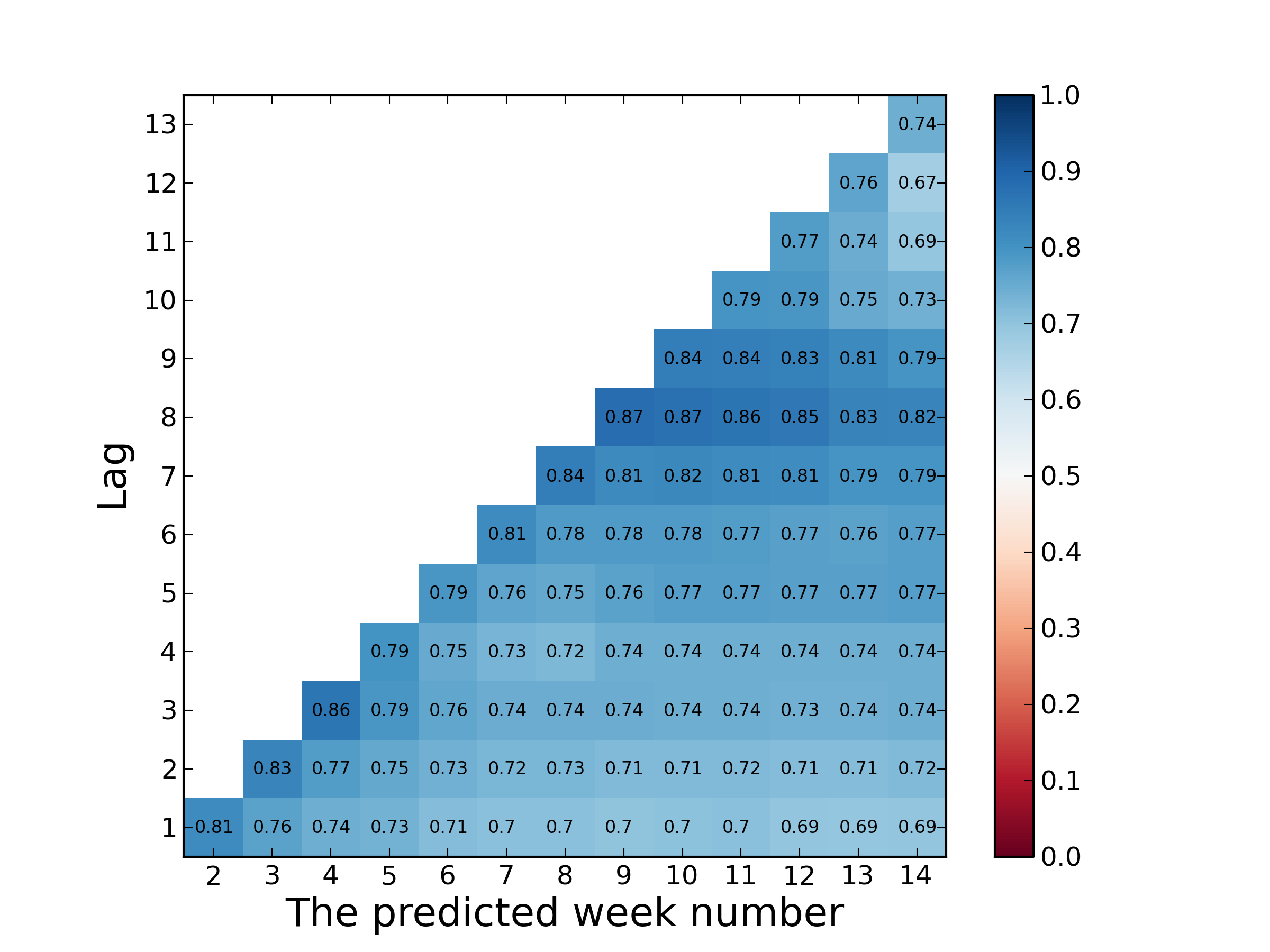}
     \caption{Logistic regression results for the \forum cohort.}\label{fig:logistic_regression_heatmap_forum_only}
\end{figure*}
\section{Predicting stopout with logistic regression}
We applied logistic regression to student persistence prediction. We used the 27 interpretive features we described earlier in this paper to form the feature vectors, and maintained the \sti value as the label.
\vspace{-5mm}
\subsection{Experimental setup}
To perform logistic regression analysis, we executed the ensuing steps for every lead, \lag and cohort combination
\footnote{
We used the logistic regression implementation of an open source machine learning library, called scikit-learn. We chose this library because it is well known and tested, fast (the core maximum likelihood estimation algorithm is written in C), with an easy to use python interface. In addition, the scikit-learn library includes an easy interface for cross validation and feature normalization.}:
\vspace{-4mm}
\begin{enumerate}
\item Performed 10 fold cross validation on the training set. As outlined in the evaluation chapter, this involved training the model on 9 folds of the train dataset and testing on the last fold.
\vspace{-2mm}
\item Trained a logistic regression model on the entire train dataset.
\vspace{-2mm}
\item Applied the model to the test dataset by putting each data point through the model then applying the decision rule in ~\ref{eq:decRule} and following the steps in ~\ref{steps} to determine the AUC under the ROC.
\vspace{-2mm}
\item Evaluating the model using mean cross validation ROC AUC and test set ROC AUC.
\end{enumerate}

\subsection{Experimental Results}\label{sect:results}
\begin{figure*}[ht!]
  \centering
    \includegraphics[width=0.8\textwidth]{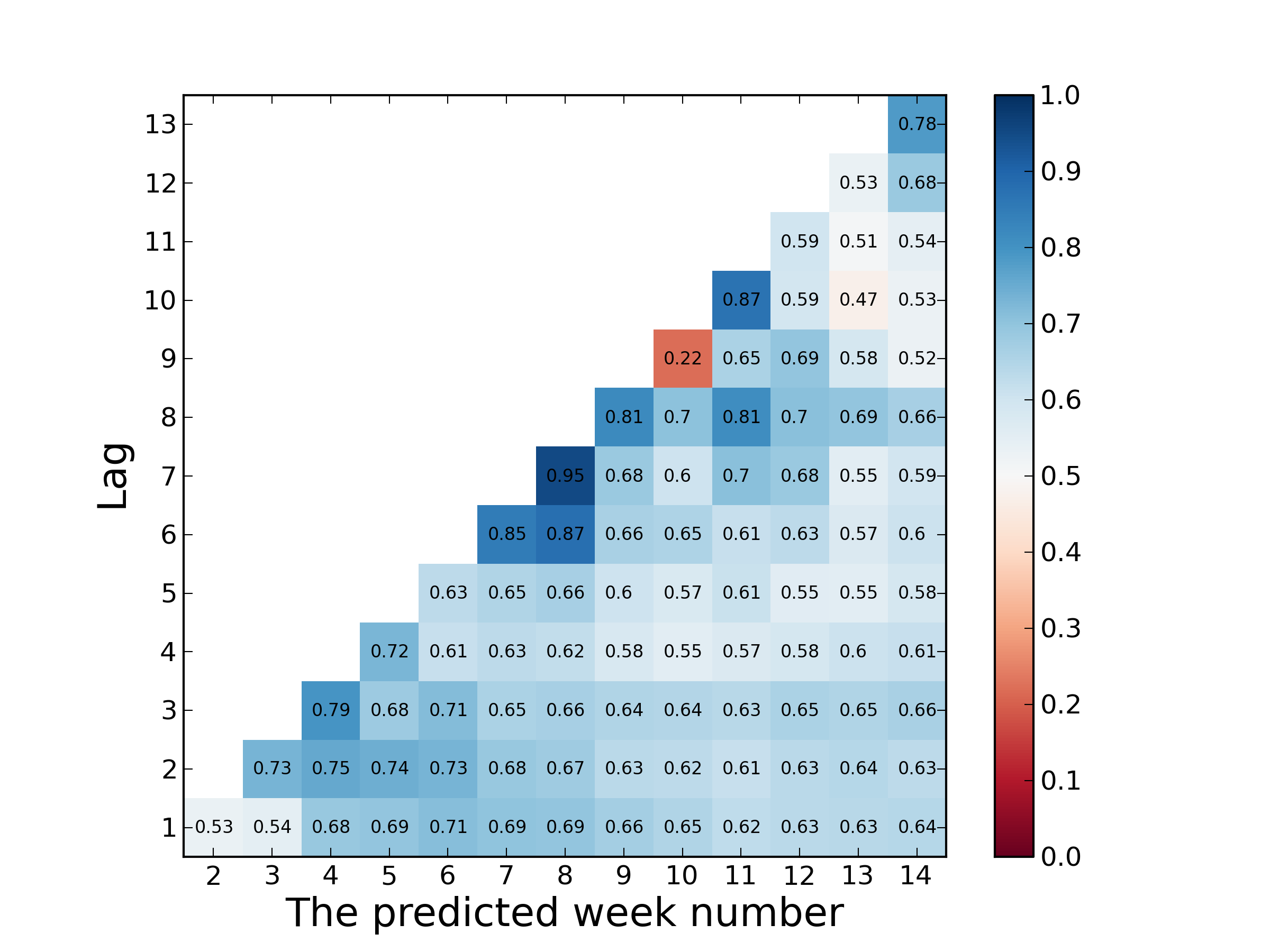}
    \caption{Logistic regression results for the \both cohort.}\label{fig:logistic_regression_heatmap_forum_and_wiki}
\end{figure*}

\begin{figure*}[ht!]
   \centering
    \includegraphics[width=0.8\textwidth]{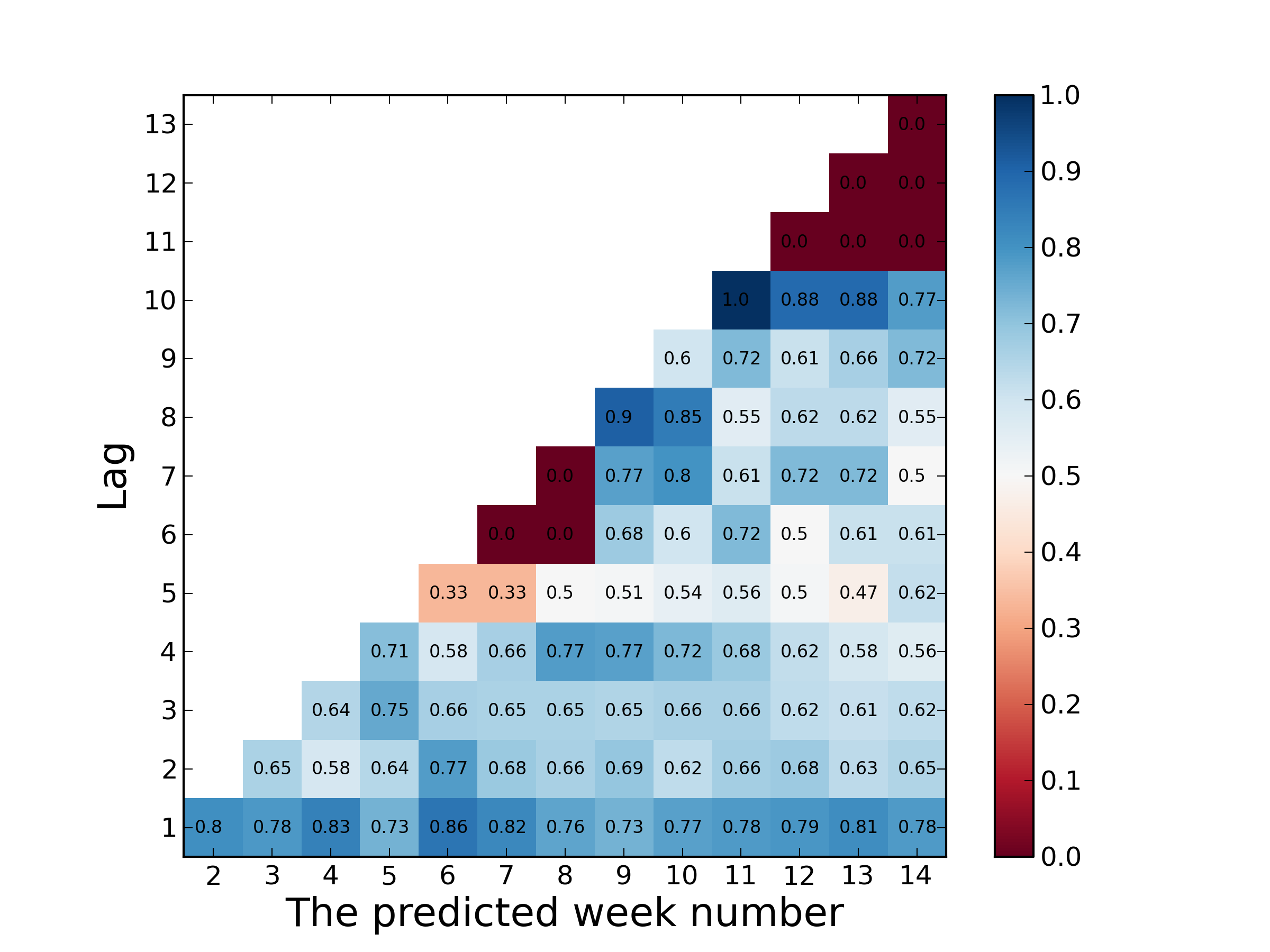}
     \caption{Logistic regression results for the \wiki cohort.}\label{fig:logistic_regression_heatmap_wiki_only}
\end{figure*}

Figures \ref{fig:logistic_regression_heatmap_no_collab} through \ref{fig:logistic_regression_heatmap_wiki_only} summarize the AUC of the receiver operating characteristic for all four cohorts over each lead and \lag combination. Overall, logistic regression predicted dropout with very high accuracy. Some experiments, such as a \lag of 7, predicting week 8 in the \both cohort achieved accuracies as high as 0.95, a noteworthy result(Figure \ref{fig:logistic_regression_heatmap_forum_and_wiki}). Moreover, the entire diagonal of the \neither cohort's heatmap (Figure \ref{fig:logistic_regression_heatmap_no_collab}) resulted in an AUC greater than 0.88. This diagonal represents experiments with a lead of one. Thus, we can surmise that the extracted features are highly capable of predicting stopout, especially when the prediction week is fairly near the \lag week.

Across all experiments, the predictive models of the \neither cohort achieved the highest predictive accuracies. This is because \neither is by far the largest cohort, which resulted in high performing, stable accuracy for all 91 experiments. Conversely, the \wiki cohort performed terribly for many experiments. In fact, for some \lag and predicted week combinations, the model could not even compute an AUC because there were not enough examples to test on.

What follows is a deeper explanation of two interesting prediction problems and their results.
\vspace{-4mm}
\paragraph{Are there early signs of stopout?}
One interesting prediction problem is trying to predict student persistence into the last week of the course using a single week of data. Practically speaking, this would enable platform providers and instructors to predict which students would finish the course by the end of the first week. Potentially, this would allow instructors to interpret the reason for student stopout as motivational (such as just browsing) rather than course-specific reasons (such as the content becoming too difficult), because the students have not been exposed to much content yet. Furthermore, early-sign stopout prediction could allow courses to target certain types of students for some type of intervention or special content. If our models are successful, the results would imply that our extracted features are capturing a student's persistence far in advance. Remarkably across cohorts, the generated models achieved an AUC of at least 0.64, and reached as high as 0.78 in the case of the \wiki cohort. 

The \wiki AUC of 0.78, or even the \neither of 0.7 suggests it is possible to roughly estimate which students will finish the course. Implications include the ability to reach out to students likely to stop the course before they become disengaged, or giving a professor a rough indication of how many students to expect each week. If these predictions hold true for other courses, a prediction model could be used to measure the success of course experiments, such as changing course content.

In the case of the \wiki cohort, the model performed well for most later predictive weeks given a \lag of one. This indicates two things. Firstly, \wiki students show remarkably strong early signs of persistence. Secondly, given more students, predictive models of the \wiki cohort would likely perform well. Owing largely to the small pool size of the \wiki cohort, model performance suffered, especially as \lag increased, because there were not enough students to appropriately train on. However, with a lead of one, the models used more student's data because we included all students who started in the course.

\begin{figure*}
        \centering
     \subfigure[\neither cohort]{\label{fig:first}
           \includegraphics[width=0.45\textwidth]{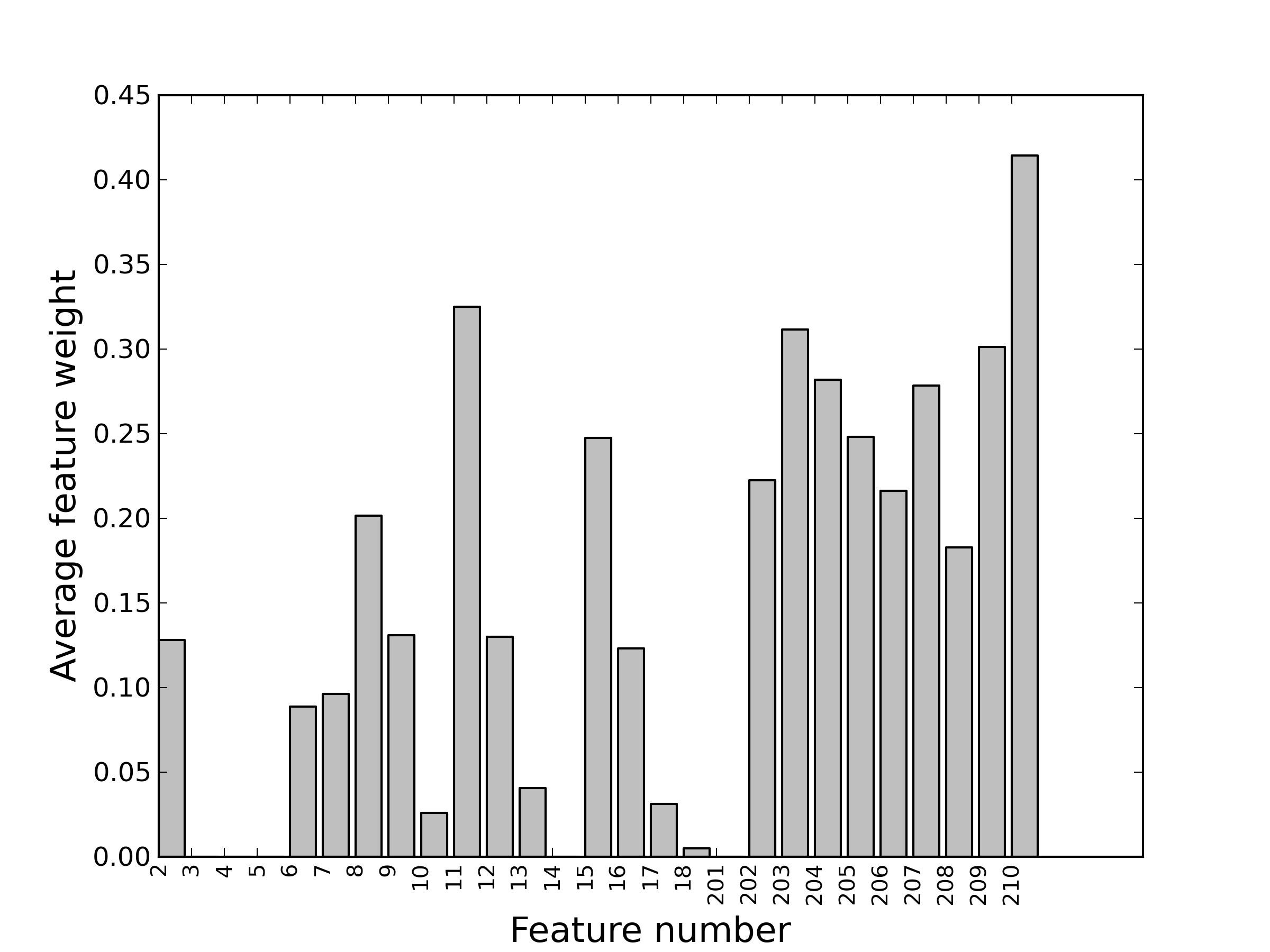}}%
        \subfigure[\forum only]{%
           \label{fig:second}
         \includegraphics[width=0.45\textwidth]{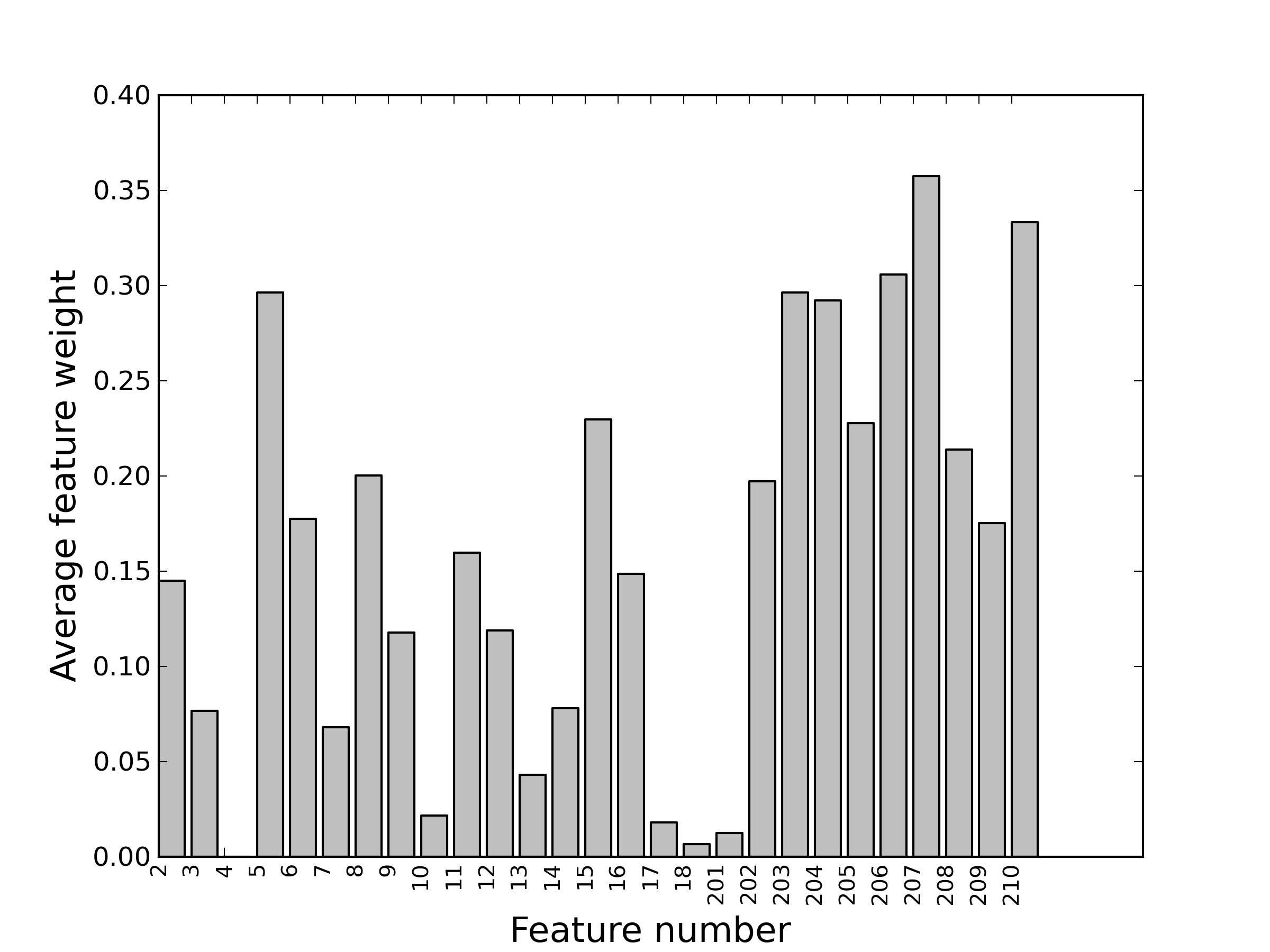}
        }\\ 
        \subfigure[\both]{%
            \label{fig:third}
           \includegraphics[width=0.45\textwidth]{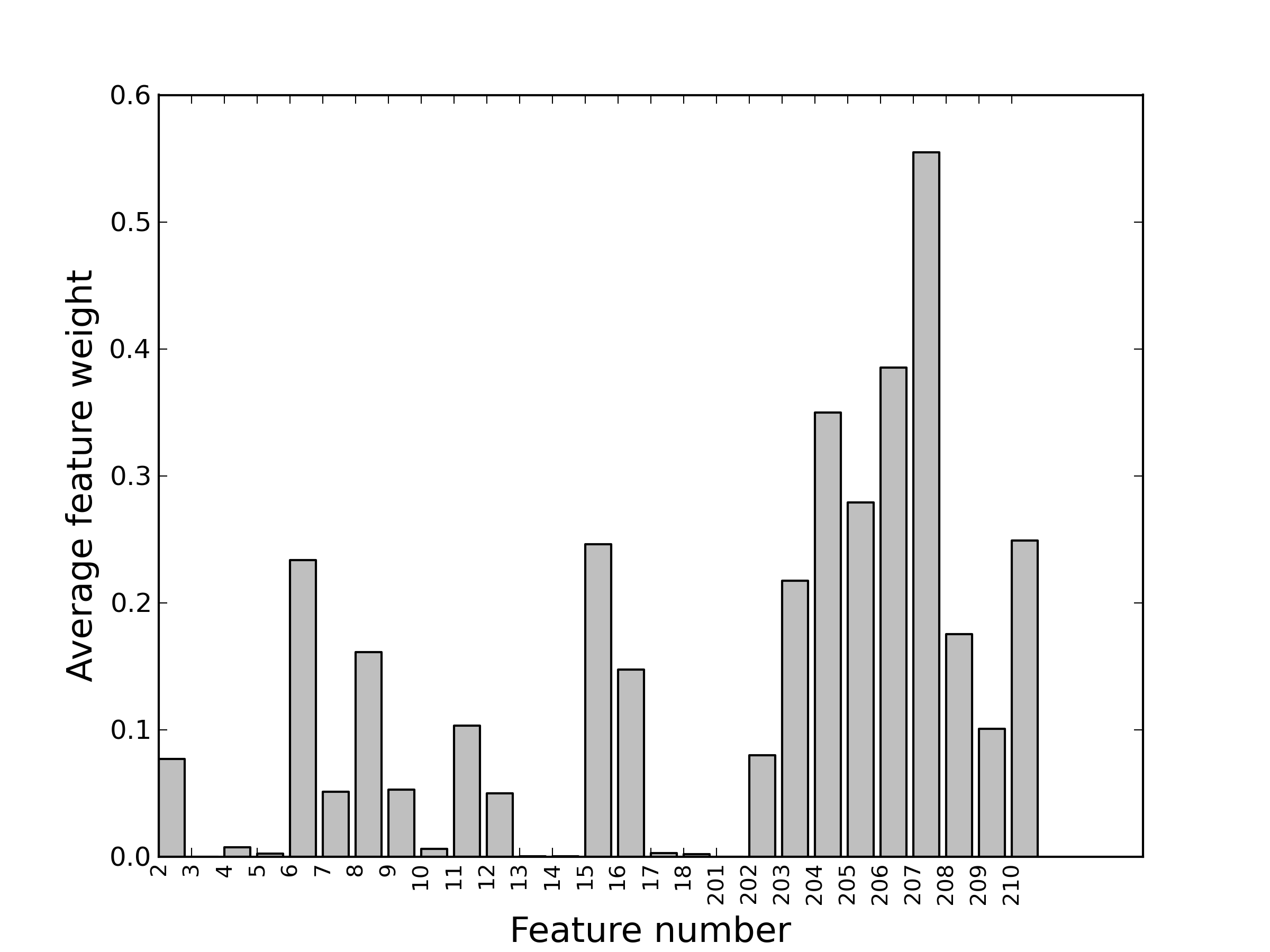}
        }
        \subfigure[\wiki only]{%
            \label{fig:fourth}
             \includegraphics[width=0.45\textwidth]{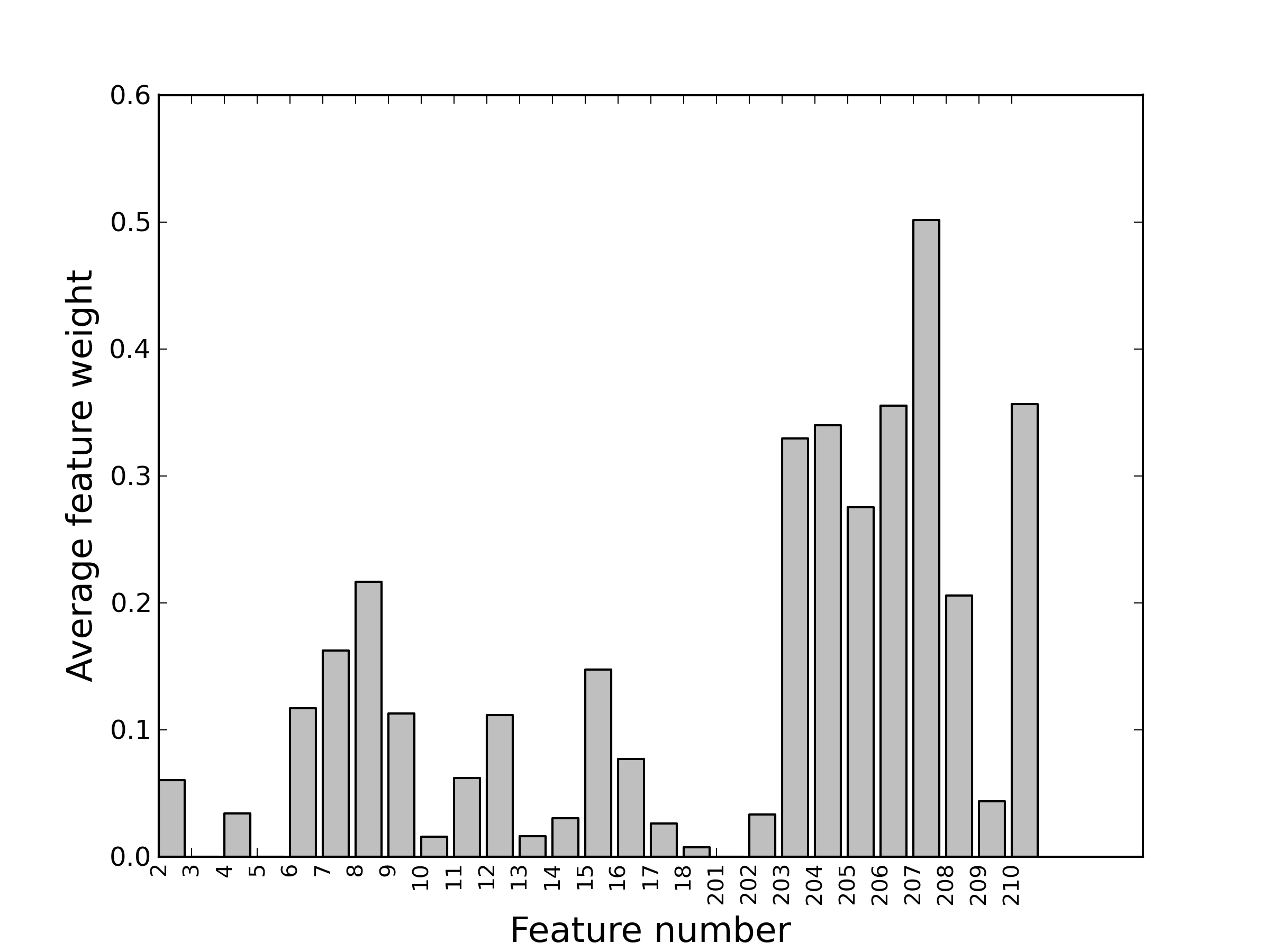}
        }

        \caption{Relative importance of different features across all variants (different lead and lag) of  stopout prediction problem. The four plots give the feature importances as found for the four cohorts we described in this paper. Summary: For the \neither cohort, top 5 features that had the most predictive power across multiple stopout predictive problems include \tten, \feleven, \tthree, \tnine, \tfive.  For the \forum cohort, top 5 features that had the most predictive power across multiple stopout predictive problems include \tseven, \tten, \ffive, \tsix, \tthree.  For the \both cohort, top 5 features that had the most predictive power across multiple stopout predictive problems include \tseven, \tsix, \tfour, \tfive, \tten. For the \wiki cohort, top 5 features that had the most predictive power across multiple stopout predictive problems include \tseven, \tsix, \tten, \tfour, \tthree. For more details about how these relative importances were calculated we refer the user to \cite{featMOOC}. }\label{fig:featImp}
\end{figure*}
\vspace{-5mm}
\paragraph{The prediction spike after the midterm}
Leading up to the midterm (in week 8), making predictions using a \lag of $i$, where $i$ is the current week, yields a fairly consistent AUC. In other words, students who will \sti after the midterm resemble their persistent counterparts up until week 8. However, using \lag 8 instead of 7, thereby including midterm data, produces an upward prediction spike in all four cohorts.

Perhaps the most striking spike example is in the most consistent cohort, the \neither students. If the model attempts to predict using a only lag 7, it realizes an AUC of 0.75. If the model expands to include midterm week data from week 8 and attempts to predict who will be in the course the next week, it achieves an AUC of 0.91. This is a significant spike. Similarly, the \both cohort increases AUC significantly from 0.68 in week 7 to 0.81 in week 8.

With the addition of the midterm week data, the model is equipped to make reasonably consistent predictions through the end of the course. In fact, for the two cohorts of significant size, the region including and beyond week 8 achieves the highest AUCs of the entire course. This suggests that the midterm exam is a significant milestone for \sti prediction. It follows that most students who complete the midterm finish the course. For the two smaller cohorts, \wiki and \both, the region beyond week 8 realizes terrible predictive power because too few students remain in the course to accurately train on. 

Consider what a wiki is (students summarize and reframe their knowledge) and the high level of engagement it reflects (more than forum). Therefore, other technologies that have similar kind of high engagement may have same influence on persistence.

\vspace{-5mm}
\paragraph{Feature importance}
{We utilized randomized logistic regression methodology to identify the relative weighting of each of the feature. More details about this approach are presented in \cite{featMOOC}. Here we briefly present the results of that experiment through Figure~\ref{fig:featImp}. In these four figures higher bar represents higher importance of that feature in predicting stopout across all 91 experiments for that cohort.  We summarize the features we found important in the findings section.}

\section{Multiple classifiers}\label{sect:mulitple_classifiers}
After successfully modeling the data using logistic regression and Randomized logistic regression, we proceeded to model the data using a number of classifiers via a cloud based machine learning as service framework called Delphi. Delphi is a first-ever shared machine learning service developed by members of ALFA group at CSAIL, MIT \cite{Will14} \footnote{http://delphi.csail.mit.edu/}.  It is a multi-algorithm, multi-parameter self-optimizing machine learning system that attempts to automatically find and generate the optimal discriminative model/classifier with optimal parameters. A hybrid Bayesian and Multi-armed Bandit optimization system searches through the high dimensional search space of models and parameters. It works in a load balanced fashion to quickly deliver results in the form of ready-to-predict models, confusion matrices, cross validation accuracy, training timings, and average prediction times. 
\subsection{Experimental setup}
In order to run our datasets through Delphi, we performed the following:
\begin{enumerate}
\item Chose a few lead, lag combinations to run on Delphi. Since Delphi creates many models, we only chose 3 datasets per cohort. We chose lead and lag combinations which were difficult for logistic regression to predict so we could see if Delphi would perform better. We chose the following combinations: lead of 13, lag of 1; lead of 3, lag of 6; lead of 6, lag of 4.
\item Flattened each cohort's train and test dataset to generate files which could be passed to Delphi. We flattened in the same manner as described in logistic regression section.
\item Ran the 12 datasets through Delphi. This gave us 12 models which performed best on a mean cross validation prediction accuracy metric.
\item Evaluated these models on the basis of test dataset ROC AUC and cross validation ROC AUC performance.
\end{enumerate}

\subsection{Delphi results}
The models created by Delphi attained AUCs very similar to those of our logistic regression and HMM models (described in \cite{hmmMOOC}. The best algorithm chosen by Delphi varied depending on which lead, lag and cohort combination was chosen. The algorithms included stochastic gradient descent, k nearest neighbors, logistic regression, support vector machines and random forests.

For the two larger cohorts, \neither and \forum, Delphi's models used logistic regression, stochastic gradient descent, support vector machines and random forests. For each of the lead and lag combinations, the models' results were within 0.02 of our logisitic regression results. This indicated that the predictive power of these cohorts was not due to the type of model used. Rather, the strong predictive accuracies achieved were due to the interpretive features in the models. As noted in \cite{featMOOC}, varying the features used, such as when using only the \selfself features rather than the \selfself features and \crowdself features, significantly changed the results. These findings lead us to conclude that focusing on better features provides more leverage in MOOC data science than does fine-tuning models.

For the smaller \wiki and \both cohorts, Delphi's models provided significantly better accuracy. For example, for the \wiki cohort, all three lead and lag combinations' models produce AUCs greater than 0.85. The best classifiers used to model these cohorts included k nearest neighbors and stochastic gradient descent. This indicates that for these cohorts, the type of model matters a great deal. We conclude that this is due to the small size of the cohorts. Some classifiers are able to more gracefully handle less data. This provides early suggestive evidence that, when a student cohort is relatively small (in relation to number of features), it is important to investigate multiple models  to identify the most accurate one.
\section{Related work and literature}\label{sect:related}
Dropout prediction and analysis of reasons for dropout have been of great interest to the educational research community in a variety of contexts: e-learning, distance education and online courses offered by community colleges. To understand and gain perspective on what has been done so far, we surveyed a large cohort of relevant literature. Tables~\ref{relatedwork} present a list of 25 research studies we surveyed that were prior to MOOCs. Table~\ref{tablemoocs} presents the list of 8 concurrent studies that correspond to MOOCs. 

We use a set of five axes along to compare research models.  For any model, the axes are 1) intended purpose: \textit{predictive} vs. \textit{correlative}, 2) whether \textit{behavioral} and/or \textit{non-behavioral} attributes were employed, 3) use of \textit{longitudinal} and/or \textit{time-invariant} variables and 4) use of \textit{trace data} and/or \textit{survey data}.  A subset of these axes are similar to those identified by \cite{lykourentzou2009dropout}.  
\vspace{-3mm}
\paragraph{Axis 1: Intended Purpose -- Predictive vs. correlative}
By categorizing a model as predictive, we identify it being used \textit{prospectively} to predict the whether or not a student \textit{will} dropout. Predictive modeling is often the basis of  interventions.  It can be used while a course is running.  In correlative modeling, analysis is performed to correlate one or more variables with completion (or progress to some timepoint). \textit{Retrospectively},  the reasons for dropout are identified. \cite{diaz2002online, tyler2006early, street2010factors} provide an excellent summary of a number of \corr studies performed in this domain. Many studies build predictive models to not operationalize the model for actual prediction during a course but to gain insights into which variables and what values of the variables are predictive of dropout. We categorize such approaches as \corr as well. 

Within the predictive models category, in the literature, there is an abundance of modeling problems set up to use a  set of variables recorded over a single historical interval, e.g first 3 weeks of the course, and to predict an event at a single timepoint. For example, using data or surveys collected from the first 4 modules of a course to forecast stopout after the midterm. In some cases, however, when predictive models are built for a number of time points in the course as in \cite{lykourentzou2009dropout} the model is not built to \textit{predict ahead}. 

In contrast, we identify 91 different predictive modeling problems within a single MOOC. We take pains to not include any variable that would arise at or after the time point of our predictions, i.e.  beyond the lag interval.  We do this so we can understand the impact of different timespans of historical information on predicting at different time intervals forward. In other words, our study is the first to the best of our knowledge to systematically define multiple prediction problems so predictions could be made during every week of the course, each week to the end of the course. \cite{lykourentzou2009dropout} provide an excellent summary of studies that fall in \pred category. 

Finally, we are concerned with the accuracy of predictive model so that it could be used during the course for intervention. As predictions can make two types of errors: \textit{mispredict} stopout or \textit{mispredict} persistence, our use of area under the receiver operating characteristic curve (AUC) as a metric for measuring the efficacy of our models rather than $R^2$ metric is a testament to the effect. The metric emphasizes the importance of both the errors and we aim at optimizing this metric. To the best of our knowledge this metric has not been used to evaluate the models. Additionally, we provide a \textit{probability of stopout} allowing the user to choose a threshold to make prediction. This allows the intervention designer to choose a trade-off point on the receiver operating characteristic curve. 

\vspace{-3mm}
%
\paragraph{Axis 2: Behavioral and/or non-behavioral attributes.}
This categorization identifies whether or not variables that capture learning behavior of student were used in modeling.  Examples of a non-behavioral attribute are a student's age, sex or location, occupation, financial status \cite{parker1999study,willging2009factors}. Second kind of non-behavioral variables are \textit{perceptual} variables, such as those derived from questionnaires, that need to be self reported. Our models do not depend on perceptual variables,  neither do they depend on non-behavioral variables such as age, gender and others. While such variables can play a significant role in increasing accuracy of models (especially when predicting far ahead), in a MOOC they may not be available. This is a powerful and significant difference as this allows us to be able to transfer the model without needing personally identifiable information.

Within the use of behavioral data, the most common behavioral variables used are performance related either prior to or during the course. For example, \cite{lykourentzou2009dropout} use prior academic performance (education level), other even use high school GPA, college GPA, freshmen year GPA \cite{morris2005predicting,mendez2008factors}. Some studies compose variables based on project grade and test grades during the course \cite{lykourentzou2009dropout}. In almost all cases, prior academic performance has been found to be the highest predictor of the student persistence \cite{mendez2008factors,xenos2002survey,allen2008prediction}. 

The second type of behavioral variables are based on students' interaction with the educational resources (online or otherwise) rather than performance on a test or a midterm, or prior academic performance, for example \textit{how much time a student spent on lecture videos} or \textit{whether or not student attended the orientation session}. We tackle the challenge of identifying variables that capture students'  interactions with the online platform using actual trace analysis (log analysis). We argue that such analysis can enable identification of attributes of the course that could be associated with stopout. For example, a difficult concept, or a rather hard/confusing video.To the best of our knowledge, the exploitation of very detailed trace oriented variables like we derive appears to be not fully exploited.

\vspace{-3mm}
\paragraph{Axis 3: Time varying vs. time-invariant variables.}
A time varying variable captures a quantity at different points in time. Many time varying variables are summaries, such as average downloads per day, total minutes watching videos per module. Sometimes they are first processed with scoring, e.g. see the engagement scoring of \cite{poellhuber2008effect}, or ranking such as the decile of a participation level each week. In contrast, a time invariant variable is constant over time, e.g. ethnicity. The important choice between these two types is whether dynamics are factored into modeling. For example, we choose time varying variables as a means of capturing behavioral trends. 

Most studies that we surveyed capture variables that are summaries over time. Some of these variables by definition are not time dependent such as \textit{attendance at class orientation} \cite{wojciechowski2005individual} and some are usually aggregated for a period in the course (or entire course): \textit{number of emails to the instructor} \cite{nistor2010participation}. In our work we operationalize variables at multiple time points in the course. In this aspect, perhaps the closest approach to ours is \cite{lykourentzou2009dropout} where authors form the time varying variables at different points of the course - different sections of the course. 

\vspace{-3mm}
\paragraph{Axis 4:  Trace  or survey data use}  Surveys play an important role in analyzing the factors related to persistence. 

Surveys allow perceptual data to be self reported and collected via questionnaires.   They permit very specific theory, such as that underlying motivation or engagement to be articulated and used as a reference point for describing a student.  They also permit the theory to be tested. Very common among these are studies that focus on collecting information about students \textit{``locus of control}" \cite{levy2007comparing, parker1999study, morris2005predicting}, \textit{satisfaction} \cite{park2009factors, levy2007comparing}, or \textit{perception of family support} \cite{park2009factors}, among others.

Many studies struggle to collect this type of the data. To overcome mistakes in manual data entry most surveys are now provided electronically. However, in many cases, not all students submit responses to surveys and questionnaires. Survey data may ask a question which a respondent fails unintentionally to answer accurately (or worse, intentionally). 
 
Trace data is typically logs and counters. It may include participation records.  In MOOCs trace data is available at a very fine grained level. Largely, it can be considered as a set of silent, passive observations. However, one needs to build interpretations for the trace data that does not directly capture student states such as \textit{attention}, \textit{motivation}, \textit{satisfaction}. 

\setlength{\arrayrulewidth}{0.5pt}
\setlength{\tabcolsep}{10pt}
\renewcommand{\arraystretch}{1.5}

\setlength{\threecoltabwid}{\textwidth - \tabcolsep * 2 * 3}

\begin{table*}[htp]
\centering
\begin{threeparttable}
 \caption{Related Literature}\label{relatedwork}
\medskip
  \begin{tabular*}{\textwidth}{ >{\raggedright\arraybackslash}p{0.40\threecoltabwid} >{\raggedright\arraybackslash}p{0.2\threecoltabwid} >{\raggedright\arraybackslash}p{0.7\threecoltabwid}}
\hlinex
Paper & \textbf{Sample size}  & \textbf{Category}  \\ \hlinex
\cite{parker1999study}  & 100 &  \corr , \beha , \timeiv \\ \hline 
\cite{xenos2002survey}&1230& \corr , \beha , \timeiv\\ \hline
\cite{kotsiantis2003preventing}&354 & \pred , \beha , \timeiv \\ \hline
\cite{xenos2004prediction} &800 &  \corr , \beha , \timeiv \\ \hline 
\cite{zhang2004identifying}&57549 & \corr \nbeha , \timeiv \\ \hline
\cite{dupin2004pre}&464 & \corr , \beha , \timeiv\\ \hline
\cite{wojciechowski2005individual} & 179& \corr , \beha , \timeiv \\ \hline 
\cite{morris2005predicting} & 211 & \pred , \beha , \timeiv  \\ \hlinex
\cite{herzog2006estimating}&23,475& \pred , \beha , \timeiv\\ \hline
\cite{levy2007comparing}&133 & \corr , \nbeha , \timeiv  \\ \hline
\cite{holder2007investigation}&259 & \corr , \nbeha , \timeiv \\ \hline
\cite{cocea2007cross}&11 & \pred , \beha , \timeiv \\ \hline
\cite{mendez2008factors}&2232 & \pred , \beha , \timeiv \\ \hline
\cite{hung2008revealing}&98  & \pred , \beha , \timev \\ \hline
\cite{moseley2008predicting}&528& \pred , \beha , \timev \\ \hline
\cite{juan2008developing}&~50 & \corr , \beha , \timev \\ \hline
\cite{boon2008risk}&1050 & \corr , \beha , \timeiv\\ \hline
\cite{aragon2008factors}  &305& \corr , \nbeha , \timeiv  \\ \hline
\cite{allen2008prediction}&50,000& \corr , \beha , \timeiv \\ \hlinex
\cite{lykourentzou2009dropout}\tnote{1}  &193&  \pred , \beha , \timev  \\ \hline
\cite{willging2009factors} & 83& \pred , \nbeha , \timeiv \\ \hline 
\cite{park2009factors}  & 147& \corr , \nbeha , \timeiv \\ \hline 
\cite{nistor2010participation}  &209& \pred , \beha , \timev \\ \hline 

\end{tabular*}
 \medskip
\begin{tablenotes}
\footnotesize
\item[1] This article contains a comprehensive overview similar to ours about a variety of studies conducted in dropout prediction over a number of years in the e-learning/online learning context. We follow some of their findings about related work and summarize in this table in addition to a few more studies we found. 
\end{tablenotes}
\end{threeparttable}
\end{table*}

\begin{table*}[htp]
\centering
\begin{threeparttable}
 \caption{Studies about student persistence in MOOCs}\label{tablemoocs}
\medskip
  \begin{tabular*}{\textwidth}{ >{\raggedright\arraybackslash}p{0.40\threecoltabwid} >{\raggedright\arraybackslash}p{0.7\threecoltabwid} >{\raggedright\arraybackslash}p{0.02\threecoltabwid}}
\hlinex
Paper & \textbf{Category}  \\ \hlinex
\cite{deboer2013bringing}&\corr , \nbeha , \timeiv \\ \hlinex
\cite{yang2013turn}  &  \corr , \beha , \timev \\ \hline
\cite{breslow2013studying} &\corr , \beha , \timeiv \\ \hline
\cite{deboer2014changing}& \corr , \beha , \timeiv \\ \hline
\cite{deboer2014tracking}&\corr , \beha , \timev \\ \hline
\cite{halawadropout}&\pred , \beha , \timev \\ \hline
\cite{ramesh2013modeling} &\pred , \beha , \timev \\ \hline
\cite{balakrishnan2013predicting}& \pred , \beha , \timev \\ \hline
 \end{tabular*}
 \medskip
\end{threeparttable}
\end{table*}
\vspace{-5mm}
\paragraph{Student persistence studies in MOOCs}
In the context of MOOCs, study of factors relating to persistence has been of great interest due to non-completion rates. There have been at least 5 correlative studies which we present in Table~\ref{tablemoocs}. These include \cite{PoellhuberMRIReport, deboer2014changing,breslow2013studying}, for more see Table~\ref{tablemoocs}.  We categorize these studies as \corr as their goal primarily is to identify variable influences on achievement or persistence. 

Research studies performed on the same data as ours in this paper show a steady progression in how variables are assembled and progress is made on this data. \cite{breslow2013studying} identify the sources of data in MOOCs and discuss the influences of different factors on persistence and achievement. \cite{deboer2013bringing} identifies the demographic and background information about students that is related to performance. \cite{deboer2014changing} assembles 20 different variables that capture aggregate student behavior for the entire course.  \cite{deboer2014tracking} posits variables on a per week basis and correlates with achievement, thus forming a basis for longitudinal study.  Our work, takes a leap forward and forms complex longitudinal variables on a \textit{per student - per week basis}. Later, we attribute the success of our predictive models to the formation of the variables. 

In \cite{PoellhuberMRIReport} a logistic regression  model with $90\%$ accuracy was (retrospectively) developed for a French language economics course delivered through the Edulib initiative\footnote{http://edulib.hec.ca} of HEC Montreal during the spring 2012 semester.  We designate this as a correlative model because completion of the final exam was used as an explanatory variable. Univariate models were first constructed to provide information on variable significance. The final logistical regression model integrated significant variables and identified behavioral engagement measures as strongly related to persistence. 

Three predictive studies closer to our study here are \cite{halawadropout, balakrishnan2013predicting,ramesh2013modeling, ramesh:aaai14}.  All three attempt to predict one week ahead (lead =1) \footnote{Except for \cite{ramesh:aaai14} which attempts to predict at three different time points in the course}. Among the papers we surveyed \cite{balakrishnan2013predicting, ramesh2013modeling} use area under the curve (AUC) as a metric for evaluating the predictive model.

There are three noteworthy accomplishments  of our study when compared to these studies above. First, throughout our study we emphasize on variable/feature engineering from the click stream data and thus generate complex features that explain student behavior longitudinally \cite{featMOOC}. We attribute success of our models to these variables (more then the models themselves) as we achieve AUC in the range of 0.88-0.90 for one week ahead for the \neither cohort.  

Second we focus on forming features/variables from highly granular, frequently collected \textit{click stream} data which allows us to make predictions for a significantly large portion of students who \textit{do not} participate in forums, and in addition captures learners interaction with resources and assignments. In the course data we worked with, only 8301 out of 52,939 students participated on forums (approximately 15.6\%, See Figure~\ref{fig:cohortsplit}).  We argue that variables derived from learner interactions on forums, as presented in \cite{ramesh2013modeling, ramesh:aaai14,yang2013turn} will only be available for a subset of learners. 

Third,  we split the learner population into four different cohorts and our methodology generates 91 different prediction problems based on different leads and lags and builds models for each of them. These result in 364 different prediction problems requiring modeling.    
\section{Summary of Research Findings}\label{sect:findings}
Our modeling and feature engineering efforts reveal the following: \footnote{While we refer the reader to \cite{featMOOC} in the compendium for detailed descriptions of the features we employed for prediction, numbered $x_1 \dots x_{18}$ and $x_{201} \dots x_{208}$, we present a summary of our findings.}:

\begin{itemize}
\item Stopout prediction is a tractable problem. Our models achieved an AUC (receiver operating characteristic area-under-the-curve) as high as 0.95 (and generally $\sim$0.88) when predicting one week in advance. Even with more difficult prediction problems, such as predicting student stopout at the end of the course with only one week's data, our models attained AUCs of $\sim$0.7. This suggests that early predictors of stopout exist.

\item For almost every prediction week, our models find only the most recent four weeks of data predictive.

\item Taking the extra effort to extract complex predictive features that require relative comparison or temporal trends, rather than employing more direct covariates of behavior, or \underline{even trying multiple modeling techniques}, is the most important contributor to successful stopout prediction. While we constructed many models with a variety of techniques, we found consistent accuracy arising \textbf{across techniques} which was dependent on the features we used. Using more informative features yielded superior accuracy that was consistent across modeling techniques. Very seldom did the modeling technique itself make a difference. A significant exception to this is when the model only has a small number of students (for example, approximately less than 400) to learn from. Some models perform notably better than others on less data.

\item A crowd familiar with MOOCs is capable of proposing sophisticated features which are highly predictive. The features brainstormed by our crowd-sourcing efforts were actually more useful than those we thought of independently. Additionally, the crowd is very willing to participate in MOOC research. These observations suggest the education-informed crowd is a realistic source of modeling assistance and more efforts should be made to engage it. See \cite{featMOOC} for more details. 

\item Overall, features which incorporate student problem submission engagement are the most predictive of stopout. As our prediction problem defined stopout using problem submissions, this result is not particularly surprising. however submission engagement is an arguably good definition.

\item In general, complex, sophisticated features, such the percentile of a student when compared to other students (\x{202}, Table \ref{table:crowd_proposed_self_extracted}), which relates students to peers, and lab grade over time(\x{207}, Table \ref{table:crowd_proposed_self_extracted}), which has a temporal trend, are more predictive than simple features, such a count of submissions (\x{7}, Table \ref{table:self_proposed_self_extracted}). 

\item Features involving inter-student collaboration, such as the class forum and Wiki, can be useful in stopout prediction. It is likely that the quality and content of a student's questions or knowledge are more important than strict collaboration frequency. We found that, in particular, the length of forum posts (\x{5}, Table \ref{table:self_proposed_self_extracted}) is predictive, but the number of posts (\x{3}, Table \ref{table:self_proposed_self_extracted}) and number of forum responses (\x{201}, Table \ref{table:crowd_proposed_self_extracted}) is not. The role of the collaborative mechanism (i.e. Wiki or forum) also appears to be distinctive since, in contrast to forum post length, Wiki edits have almost no predictive power.
\end{itemize}

\section{General reflections for the entire compendium}\label{sect:general}
This extensive project has revealed a combinatorial explosion of MOOC modeling choices.  There are a variety of algorithms which one could use, a variety of ways to define a modeling problem  and a number of ways to organize data which is fed into modeling.  There are also numerous challenges in assembling features while features themselves  turn out to be of very high importance. One has work systematically and be thorough in feature definition and model exploration, otherwise one will never know if one has derived the best prediction capability from the data.  

To successfully apply the power of data science and machine learning to MOOC analytics, multiple aspects of the process are critical:

\paragraph{Feature engineering}
One has to be meticulous \textbf{from the data up} -- any vague assumptions, quick and dirty data conditioning or preparation will create weak foundations for one's modeling and analyses. Many times painstaking manual labor is required - such as manually matching up \textit{pset} deadlines, etc. We need to be ready to think creatively as you brainstorm and extract features, and be flexible in the ways we assemble them. For example, utilizing the crowd is much richer than just our own expertise.

\paragraph{Machine learning/modeling at scale} 
There are many ways to represent the extracted features data- with or without PCA, temporal and non-temporal, discretized and non discretized. Additionally there are a number of modeling choices - discriminative, generative or mixed models which include many types of classifiers. One has to consider a number of them to enable insights at scale. The alternative results in a much smaller scope with more limited results.

Our ability to build 10,000 models relied on us first building the cloud scale platforms. This is especially true as the machine learning process includes iterations over data definitions, features and cohort definitions. Only through a large scale computational framework are these multiple iterations possible. Throughout our analysis we ran on hundreds of nodes simultaneously, using the DCAP and Delphi frameworks. 

\paragraph{Transfer learning prospects}
In order to have a lasting impact on MOOC data science, we have to think big! Investing resources only in investigating stopout for one course limits the impacts of the results. With this in mind, we set out to create a reusable, scalable methodology.

From the beginning of the research, we have envisioned creating open source software. This would allow other researchers to apply our methodology to their own MOOC courses. That our software can be used by any other MOOC research, is due to standardization via the shared MOOCdb data schema.  Our attention to the scalability of our methods for large data sets also supports wide applicability. The prospect of multiple studies and multi-course studies would be very exciting and most welcome.
\bibliography{mooc} 
\bibliographystyle{icml2013}

\end{document}